\documentclass[10pt,conference]{IEEEtran}
\makeatletter
\def\ps@headings{%
\def\@oddhead{\mbox{}\scriptsize\rightmark \hfil \thepage}%
\def\@evenhead{\scriptsize\thepage \hfil \leftmark\mbox{}}%
\def\@oddfoot{}%
\def\@evenfoot{}}
\makeatother
\usepackage{amsmath}
\usepackage{graphicx}
\usepackage{epsfig}
\usepackage[caption=false,font=footnotesize]{subfig}
\usepackage{amssymb,url}

\newcommand {\beq} {\begin{equation}}

\newcommand {\eeq} {\end{equation}}
\newcommand {\barr} {\begin{array}}
\newcommand {\earr} {\end{array}}
\newcommand {\bear} {\begin{eqnarray}}
\newcommand {\eear} {\end{eqnarray}}
\newcommand {\bears} {\begin{eqnarray*}}
\newcommand {\eears} {\end{eqnarray*}}

\begin{document}

\title{Modelling View-count Dynamics in YouTube}
\author{\IEEEauthorblockN{C\'{e}dric Richier\IEEEauthorrefmark{1}, Eitan Altman\IEEEauthorrefmark{2}, 
Rachid Elazouzi\IEEEauthorrefmark{1}, Tania Jimenez\IEEEauthorrefmark{1}, Georges Linares\IEEEauthorrefmark{1} 
and Yonathan Portilla\IEEEauthorrefmark{1}} 
\IEEEauthorblockA{\IEEEauthorrefmark{1}University of Avignon, 84000 Avignon, FRANCE\\ 
Email: firstname.lastname@univ-avignon.fr}
\IEEEauthorblockA{\IEEEauthorrefmark{2}INRIA, B.P 93, 06902 Sophia Antipollis Cedex, FRANCE\\
Email: eitan.altman@inria.fr}}

\maketitle
\begin{abstract}
The goal of this paper is to study the behaviour of view-count in 
YouTube. We first propose several bio-inspired models for
the evolution of the view-count of YouTube videos. We show, using
a large set of empirical data, that the view-count for
90\% of videos in YouTube can indeed be associated to at least one of these
models, with a Mean Error which does not exceed $5\%$.
We derive automatic ways of classifying the view-count curve
into one of these models and of extracting the most suitable
parameters of the model. We study empirically the impact of videos' popularity
and category on the evolution  of its view-count.
We finally use the above classification along with the automatic parameters
extraction in order to predict the evolution of videos' view-count.
\end{abstract}
\begin{IEEEkeywords}
YouTube,bio-inspired models, view-count.
\end{IEEEkeywords}

\section{Introduction}
YouTube has been one of the most successful user-generated video sharing sites since its establishment in early 2005 and 
constitutes currently the largest share  of Internet traffic. The rate of 
subscription to YouTube as well as the rate of submitted videos has been growing steadily
ranking YouTube and none of its 
competitors has achieved a similar
success \cite{comScore,  comScore1}.  An 
important aspect of videos in YouTube is 
their popularity, which is defined as the number of view-counts.  
Understanding and predicting the popularity
is useful from a twofold perspective: On one
hand, more popular content generates more traffic, so understanding popularity has a direct impact on  caching and 
replication strategy that the provider should adopt; and on the other hand, popularity has a direct economic impact. 
A number of researchers have analyzed  the popularity characteristics 
of user-generated video content for 
understanding the  processes governing their popularity dynamics  
\cite{ChaUtube,crane2008viral,Gill07youtubetraffic,RatkiewiczBurstyPoP,ChShFa10,ChaTON}, with the aim of developing 
models for early-stage prediction of future popularity \cite{SzaboPop}.    
There has been also interest in understanding what 
important factors lead  some videos to become more popular than others. 
But few works have studied the temporal aspects of the popularity dynamics using some  metrics such as  view-counts,  
ratings and number of comments \cite{ChaUtube, dale, mitra}.

In this paper we describe some of the most typical behaviour of the view-count of videos in YouTube. This allows us to 
provide in-depth analysis and develop  models that capture the key properties of the observed popularity dynamics. 
Our goal is to match observed video view-counts with
one of several dynamic models. To select  candidates for these models, we
turned to bio-inspired dynamics as
we believed that the propagation of a content in 
YouTube has a strong similarity with the 
temporal behaviour of an infectious disease, which is a classical 
topic in mathematic biology \cite{bailey, meyers}.  Such
models of diseases spread  have already been used in order 
to model the spread of viruses in computer networks  
\cite{chakrabarti, ganesh}.  They have been also  used  in marketing 
for capturing the life cycle dynamics of a new product \cite{bass}
. A 
large number of papers in marketing have shown 
that product sales life cycle follow an S-curve pattern in which the 
product sales initially grow at fast 
rate and   it falls off as the limit of the 
market share is approached \cite{mahajan}.  

We propose several information 
diffusion models to classify  a dataset of more than 
800000 videos randomly extracted from YouTube and aged 
between 5 and 2500 days.  
In particular, we  exhibit six mathematical models  to which we fit videos
in our dataset. We then propose automatic ways for associating each video to 
one of the considered models.
The first criterion in selecting the model is
related to the size of the population that may be 
potentially interested by the content.  
We differentiate between models in which the
population potentially interested in the content is
nearly  constant  (we call this the "fixed target population
property") and those in which it  grows in time
(inspired by the branching process terminology, we  call this "immigration").
The fixed target population property  occurs
 in some video categories in YouTube 
as news, sport and movies.  Indeed, videos in these categories 
reach quickly the peak of the popularity and then
within a short time the diffusion dies out and the view-count does not further 
increase.

The second criterion in the classification
concerns the structural virality.   
A model is said to be viral (or to have the viral property)
if contaminated nodes
(these are the viewers of a video) have a significant
role in the propagation of the video through sharing or
embedding. It is non-viral if the propagation of the video
essentially relies on broadcast of the video from the source
(it  is then said to have the broadcast property).
In that case, a large fraction of potential target population can receive 
the information directly from the source. 

Our contribution can be summarised in the following key points 
\begin{itemize}
\item We propose six mathematical biology-inspired models  and  we show that at least 90\% of videos in YouTube are associated to one of these
six mathematical models with a Mean Error Rate  less than 5\%.  We further
show how to extract the model parameters for each video.
 
  \item We study the robustness of these models to the different thematic
categories of the video in YouTube and to different values of the peak popularity
of the video.  We show that the fraction of videos withing a given model
is quite robust and shows little dependence on the different 
thematical categories  of the video, except for Education category
which has a different behaviour:   
For this category it seems that the word-of-mouth is the dominate  mechanism through which contents are  disseminated.  
The bio-inspired models that we 
selected are further shown to be robust with respect to the peak popularity
of the video but the distribution among them is slightly different between
those videos that have acquired less than 1000 views and the rest of the 
videos.

 \item In more than 80\% of videos in YouTube, the potential population 
interested in the video increases over time.  

\item Two of the six models (The modified negative exponential   and 
modified Gompertz   models) 
cover most of videos in our YouTube Dataset (more than 75\%).  Both models capture the case of immigration  process in which the potential population or 
the ceiling value become dynamic.   
Further, the modified negative exponential   characterizes the dynamic of a non-viral content  and  it predicts 
that the accumulated number of view doesn't contribute to the propagation of the content.  
This model corresponds to the scenario wherein the content has been broadcasted to a pool of users.   
On the other side, the Gompertz model  captures viral videos  in which a part of this dynamic is propagated through word-of-mouth. 

\item
We finally use the above classification along with the automatic parameters
extraction in order to predict the evolution of videos' view-count.
We consider two scenarios: In the first we use half of the view-count 
curve as a training sequence  while  in the second one,
we take a fixed training sequence that corresponds to the first 50 days 
in the lifetime of the video. We then
compare the predicted curve to the actual one and study the prediction capacity 
within a given error bound.

\end{itemize}

\section{Setting and data }\label{data_how}
Since we intend to study different types of dynamic evolution of the view-count in YouTube, we need to collect a huge number of videos which  are available to the  general public.   

%


In this section we describe how we collected the dataset used in this study. On YouTube, a video is accompanied by a set of valuable data as title, upload time, view-count, 
related videos. The video web page also provides some statistics which are available if the content's owner allows it.

YouTube provides two APIs which allow to retrieve some of those data : the YouTube Data API for collecting static data (which are available for every user) and the 
YouTube Analytics API for seeking video statistics such as dynamics of a content (which are only available for content's owner).  Since some data cannot be collected 
through the APIs, we used a tool named YOUStatAnalyzer~\cite{YOUStatAnalyzer} in order to collect all valuable data. 


The collected data are stored in a noSQL database (MongoDB).
The noSQL solution has been chosen to allow dynamic insertion of new features for future works.
The dataset used for this study contains more than $80000$ videos randomly extracted from YouTube and aged between $5$ and $2500$ days. 
This dataset contains some static information for each video such as YouTube id, title of the video, 
name of the author, duration and list of related videos.  It also provides the evolution of some metrics (shares, subscribers, watch time and views) in a daily form 
and in a cumulative form, from the upload day till the date of crawling.

%



\section{Popularity growth patterns}\label{models}
We focus the analysis on view-count as the main popularity metric of a video. Previous analyses of YouTube showed a strong correlation 
between  view-count and other metrics as number of comments, favourites and rating. Further, these metrics correlation becomes stronger with popularity \cite{ChShFa10}.  
We model the dynamic evolution of view-count some mathematical models from the biology.
We classify the evolution of view-count in 
YouTube using two criteria:  
\begin{itemize}
 \item \textbf{Size of the target population:} 
 The target population size is the maximum number of individuals that can be, potentially, interested by the content. 
A target population belongs to one of these two types: (i) a fixed finite target population 
or (ii) a potential target population that grows in time which we call \textbf{the immigration process}.
 \item \textbf{Virality:} 
 A content is viral if the population that has seen the content participates actively in the
dissemination of the content. The content spreads like a virus does in epidemics.
Thus the probability that an individual who has not seen the content so far got it by someone who see it, increases in time.
On the contrary, a non viral content is one for which former viewers scarcely alter the diffusion process.
\end{itemize}

In the following we describe the  dynamic models in biology and their uses. 
These dynamic models have been hypothesised to describe the contagion phenomena and each 
model has its own set of assumptions about how users are infected by others. These models may 
provide some answers about the behaviour of users in YouTube  even if 
this behaviour remains notoriously difficult to quantify. 
\subsection{Fixed target population}
 \subsubsection{Viral content}
 To describe  the viral content  with fixed target  population, we use  the logistic model or the Gompertz model. These models have been used in technology 
forecasting and are referred as "S-shaped" curve.  We test these models to capture the evolution of view-count of a video in YouTube  since there is a strong similarity 
between  a video posted in YouTube and a new product launched into the marketplace. Indeed, as showed in different problems in marketing, technology product is often 
growth slowly followed by rapid exponential growth  and finally  it falls off as limit of market share is approached.  
\subsubsection*{Logistic  model}\label{sigm_mod}
The  logistic model is  a common sigmoid function which  describes the evolution  of view-count  of a video with  fixed target population.  This is a  first order 
non-linear differential equation  of the  form
\begin{equation}
 \frac{dS}{dt} = \lambda S(M - S)
 \label{eq0}
\end{equation}
where $S$ is the number of view-counts of a video and $M$ is the maximum size of the (potential) population that could access the content. This is a standard equation 
in epidemiology for describing the evolution of the number of infected individuals under the assumption that all infected nodes have developed an immunity from infection 
or these infected nodes stay infected and will not be changed to uninfected state.  Hence the infection rate  is a function  of the rate $\lambda$ and the 
size of the infected population. For the YouTube case,  this model corresponds to the scenario wherein users  may watch a video one time  and the probability to watch 
it again is negligible. A solution to equation (\ref{eq0})  is given by
\[S(t) = \frac{M}{1 + (\frac{M - S(0)}{S(0)}) e^{-\lambda Mt}}\]
This function shows that initial exponential growth is followed by a period in which  growth  starts to decrease as approaching  the maximum size of population.
%

\begin{figure}[t]
 \centering
    \includegraphics[width=0.45\columnwidth]{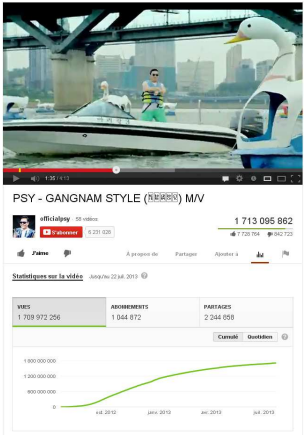}
 \caption{Example of non symmetric S-shaped behaviour of view-count in YouTube (viral).}
 \label{psy}
\end{figure}

The S-shape of the sigmoid model curve is symmetric.  But in the context of view-count, the convex phase and the concave phase could not always  
be  symmetric.  This is shown by the example in  Figure~\ref{psy}. For covering these cases we consider the Gompertz model.
\subsubsection*{Gompertz model}
\begin{figure*}[ht!]
\centering
\subfloat[Varying M \label{fig_first_case}]{
\includegraphics[width=0.3\textwidth]{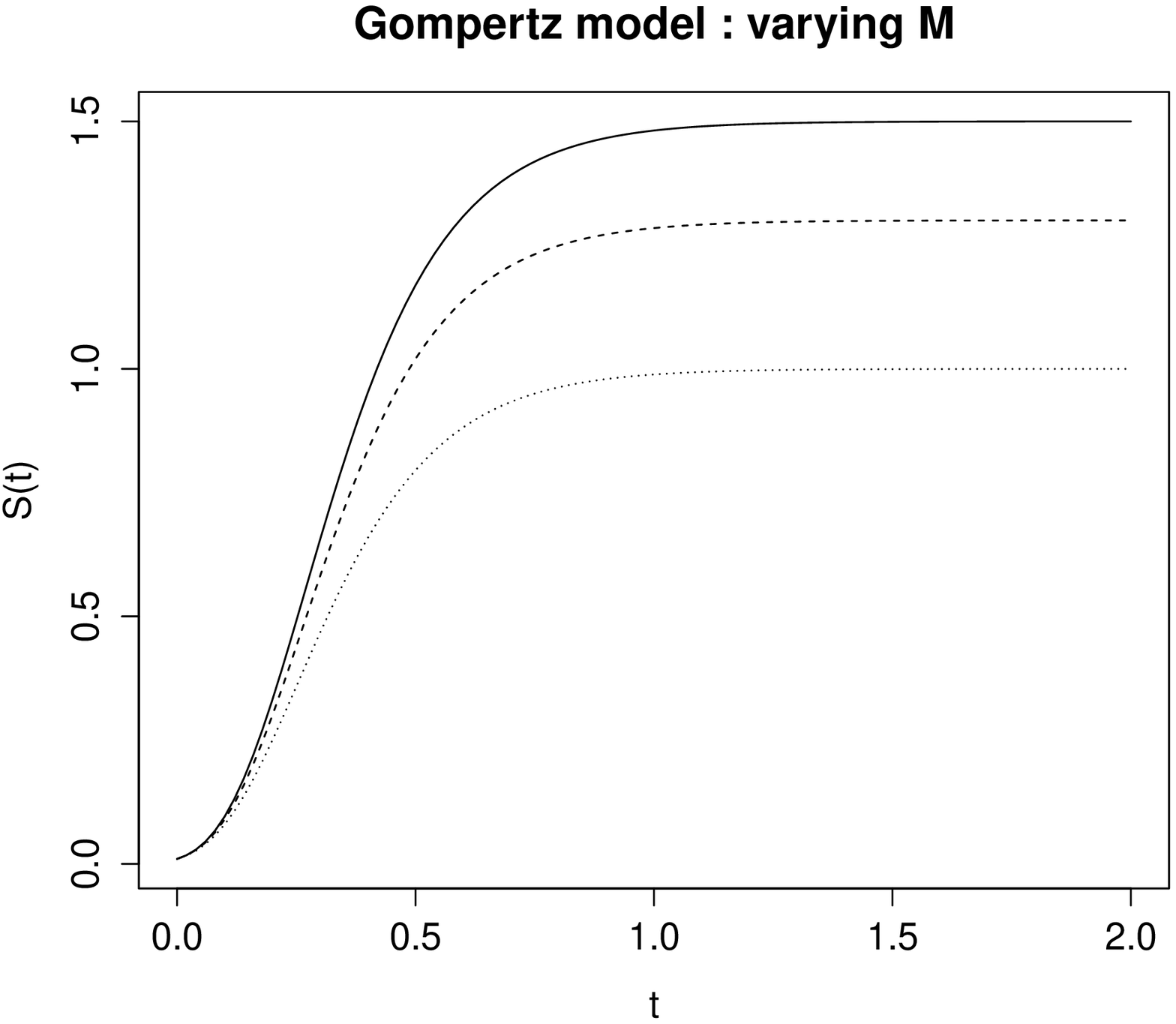}}\quad
\subfloat[Varying $\lambda$ \label{fig_second_case}]{
\includegraphics[width=0.3\textwidth]{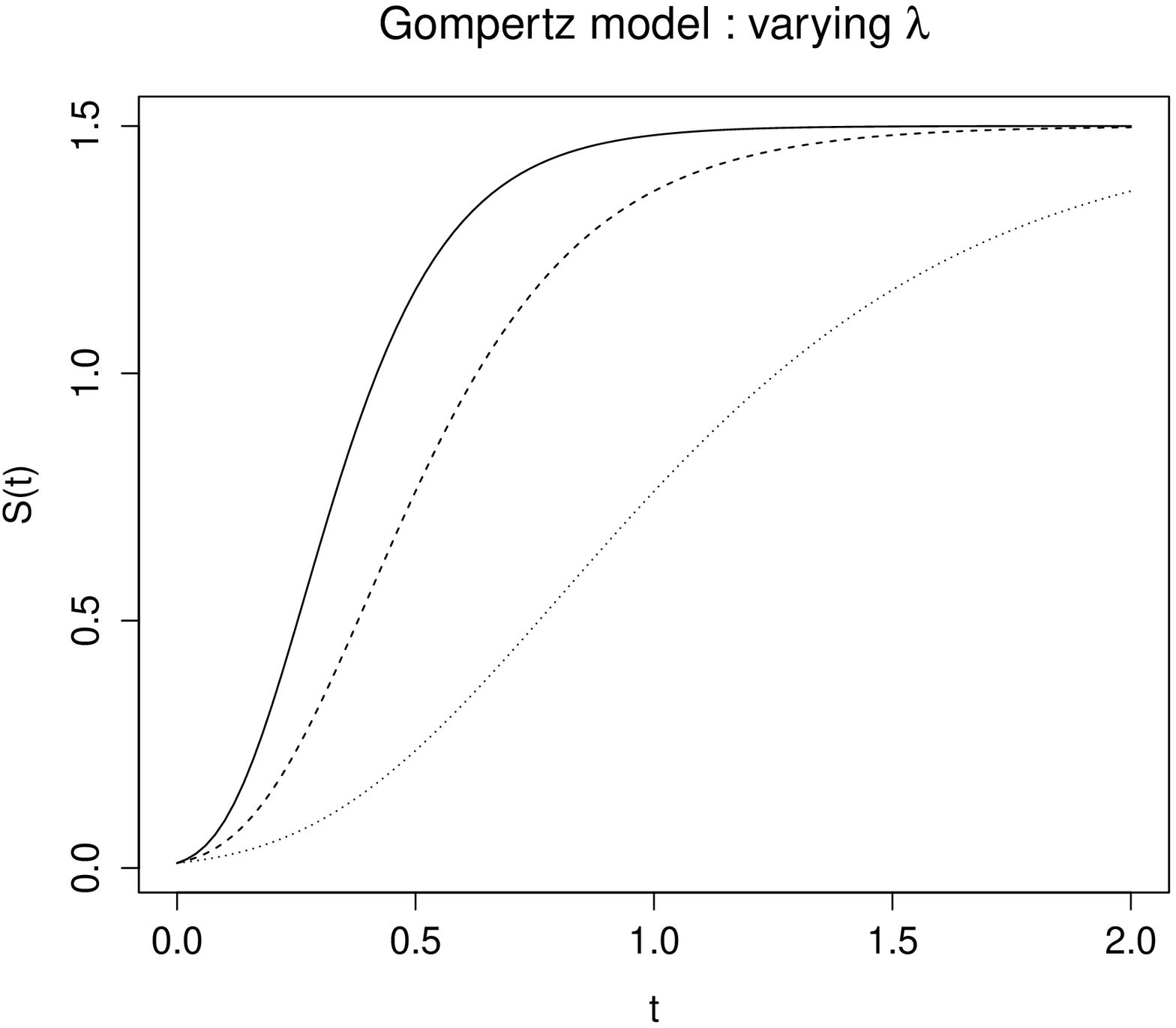}}\quad
\subfloat[Varying $S(0)$\label{fig_third_case}]{
\includegraphics[width=0.3\textwidth]{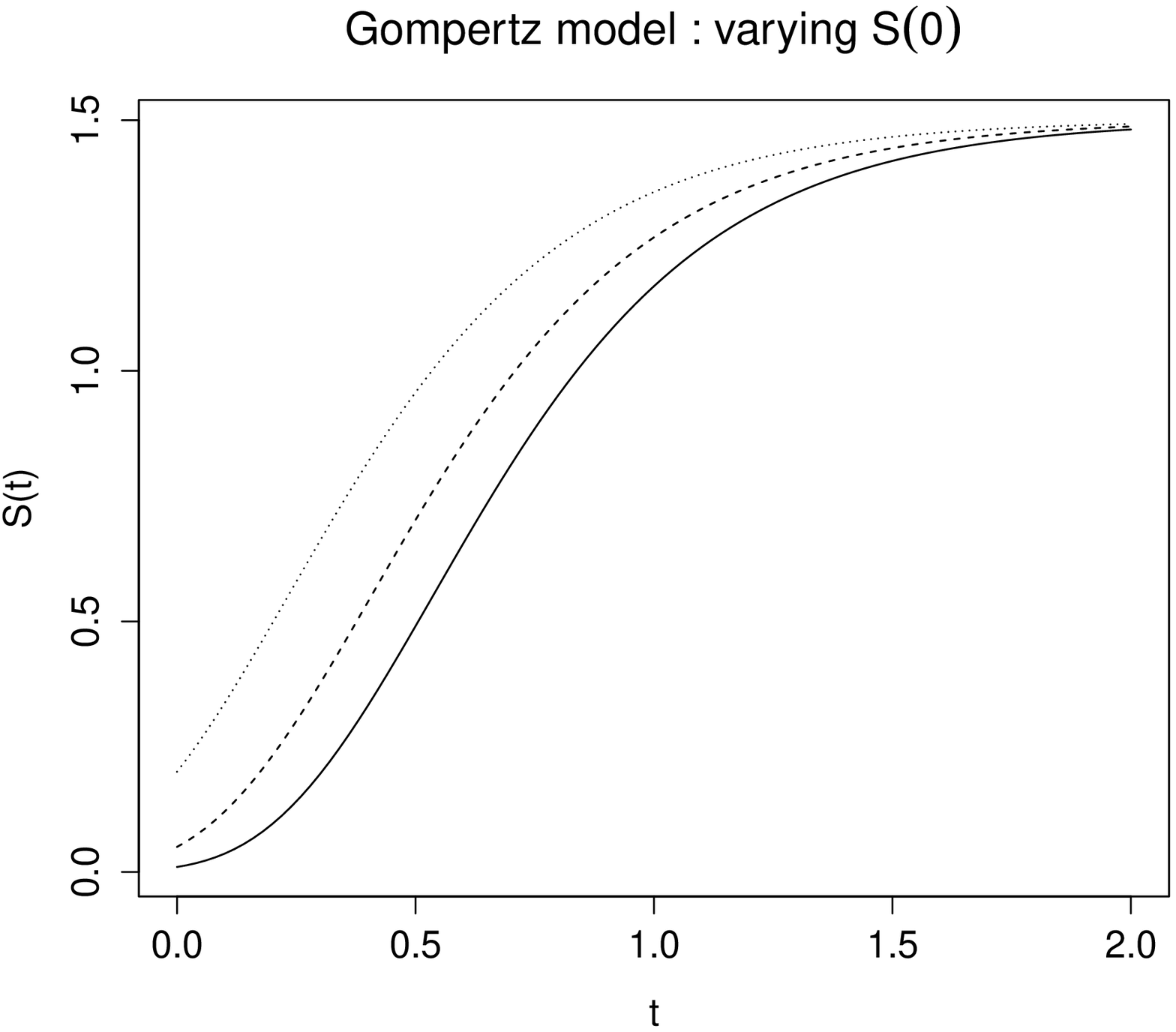}}
\caption{Graphs of Gompertz functions, showing the effect of varying one of  $M$, $\lambda$, $S(0)$ while keeping the others constant}
\label{GompParameterz}
\end{figure*}

A model which deals with the problem of symmetry of the Logistic model is given by the following dynamic equation:
\begin{equation}
 \frac{dS}{dt} = \lambda S\log(\frac{M}{S}),
 \label{gomEq}
\end{equation}
This model is called Gompertz model, and has been also used  as diffusion model of product growth.  
A solution of equation~\ref{gomEq} is given by the Gompertz function : 
\[S(t) = M\exp(-\log(\frac{M}{S(0)})\exp(-\lambda t)),\]
Figure~\ref{GompParameterz} shows  the effect   of varying one of  $M$, $\lambda$, $S(0)$ while keeping the others constant.

This model is similar to the logistic curve but it is not symmetric about the inflection. In general the Gompertz's model reaches this point early in the growth trend. 
This behaviour seems to fit well for some YouTube view-count evolution dynamics.
 
\subsubsection{Non-viral content}
A non viral content  describes the case where users do not contribute on the propagation of the content.   This is the case when the time scale of the 
content diffusion is very large compared to the size of potential population.  Hence this dynamic can  model  the case where contents gain popularity through 
advertisement and other marketing tools:  examples are when advertisement is broadcasted to a very large pool of users of a social network and people access the content 
at random thereafter.   Hence we assume that the evolution dynamic of the content follows the linear differential equation: 
\begin{equation}
 \frac{dS}{dt} = \lambda(M - S)
 \label{negExp1}
\end{equation}
The solution of  (\ref{negExp1}) is given by : 
$$S(t) = S(0) + (M - S(0))(1 - e^{-\lambda t})$$

\subsection{Growing population}
The assumption that the population is fixed, is often a reasonable approximation  when the evolution of the popularity of a content  increases quickly  and dies out 
within a short time.   But for many cases, this assumption becomes inappropriate  when the time before reaching the saturation region is longer.  

Here we consider the case of  immigration process in which the potential population growth  and the dynamic  of view-count of a content are intricacy linked.  
To capture such dependence we consider  different growth scenarios that  model the viral case and non viral case.  In this paper we restrict 
our study on the case where the target population grows with a fixed speed.

\subsubsection{Non-viral content }\label{nv_imm}
The linear growth model $S(t) = S(0) + \lambda t$ describes in a simple way the situation where users  do not 
contribute to propagate the content to other users but the content  benefits of the immigration process which gives a linear growth of  the view-count.

Another kind of non-viral curves observed are concave curves (given by the negative exponential model) which do not  converge to a flat line but become linear at the 
horizon  due to the immigration process influence. Such  dynamics could be modelled by modifying solutions of 
equation~{\ref{negExp1}} where a linear component is added:
\[S(t) = S(0) +  (M - S(0))(1 - e^{-\lambda t}) + kt\]
where $k$ is the rate of the target population growth.
Note that $S$ no more respects the equation{~\ref{negExp1}} but $(S - kt)$ does.

\subsubsection{Viral content }\label{vir_imm}
Now we consider the case when  the immigration process appears in the case of viral contents.  
In this dynamic the view-count curve  first adopts a viral behaviour (in a S-shaped phase) and then grows linearly.
%


One candidate solution to describe such a behaviour of view-count  is to add a linear  component to the Gompertz function : 
$$S(t) = M\exp(-\log(\frac{M}{S(0)})\exp(-\lambda t)) + kt$$
This dynamic seems to be relevant  according to some examples in the dataset.

\section{Dataset and Data fitting}
This section describes how we use the models presented in section~\ref{models} in order to classify the YouTube contents in our dataset.

\subsection{Dataset}\label{data_what}
\begin{figure*}[ht!]
 \centering
  \subfloat[YouTube categories distribution \label{cat_dist}]{
  \includegraphics[width=0.26\textwidth]{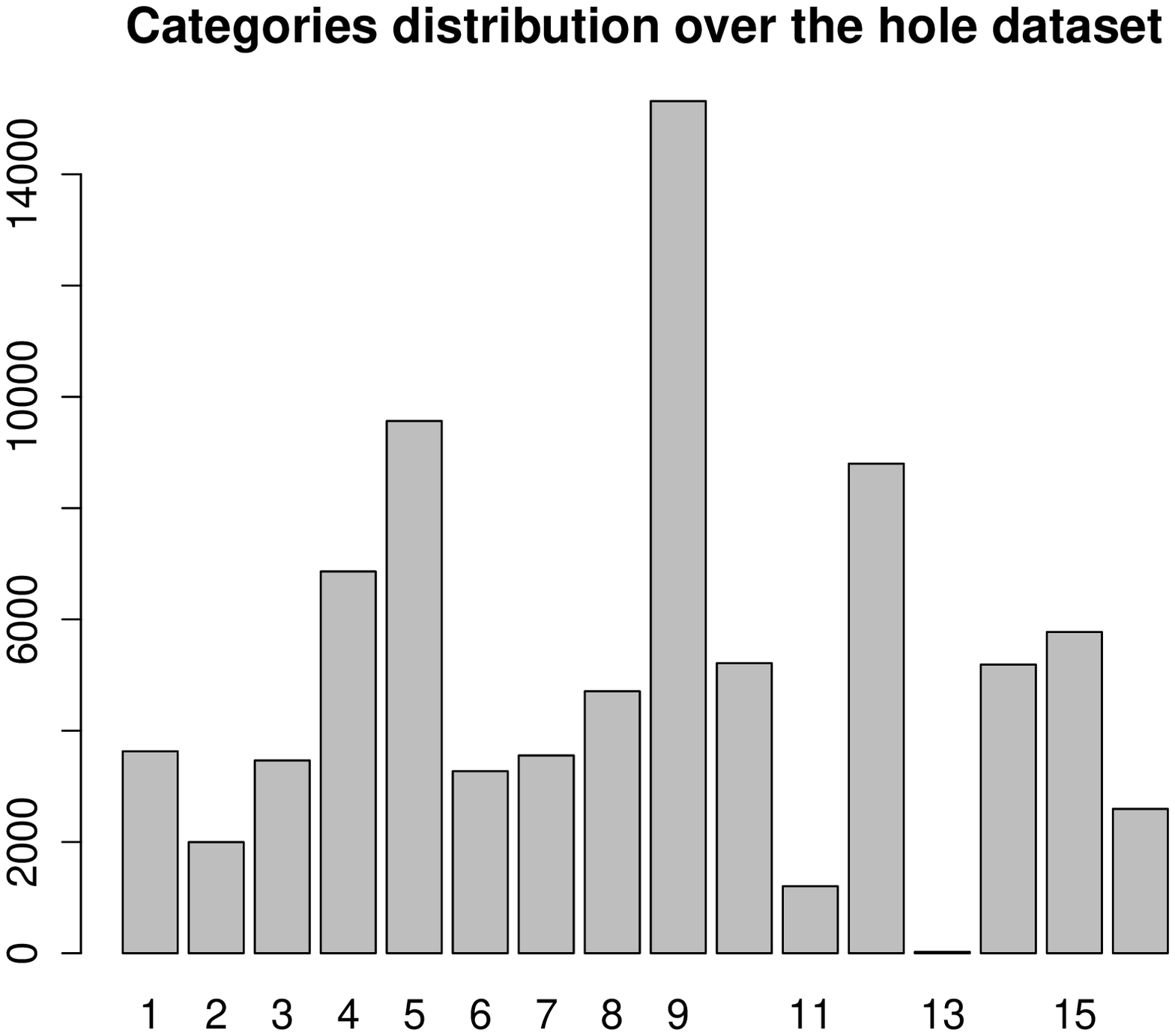}}\quad
  \subfloat[Age distribution (in number of days) \label{age_dist}]{
  \includegraphics[width=0.26\textwidth]{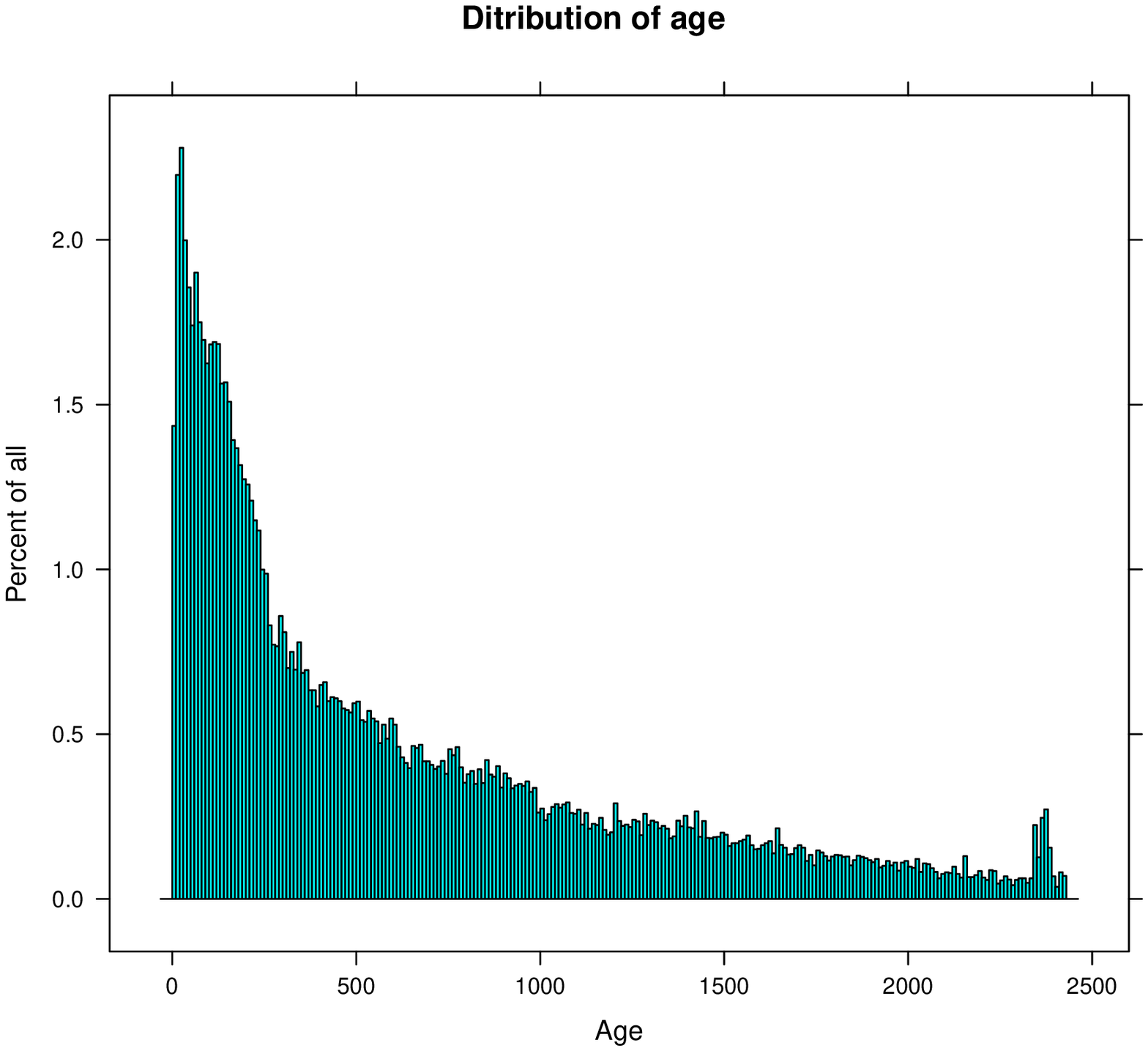}}\quad
  \subfloat[Logarithmic distribution of popularity \label{logPop_dist}]{
  \includegraphics[width=0.26\textwidth]{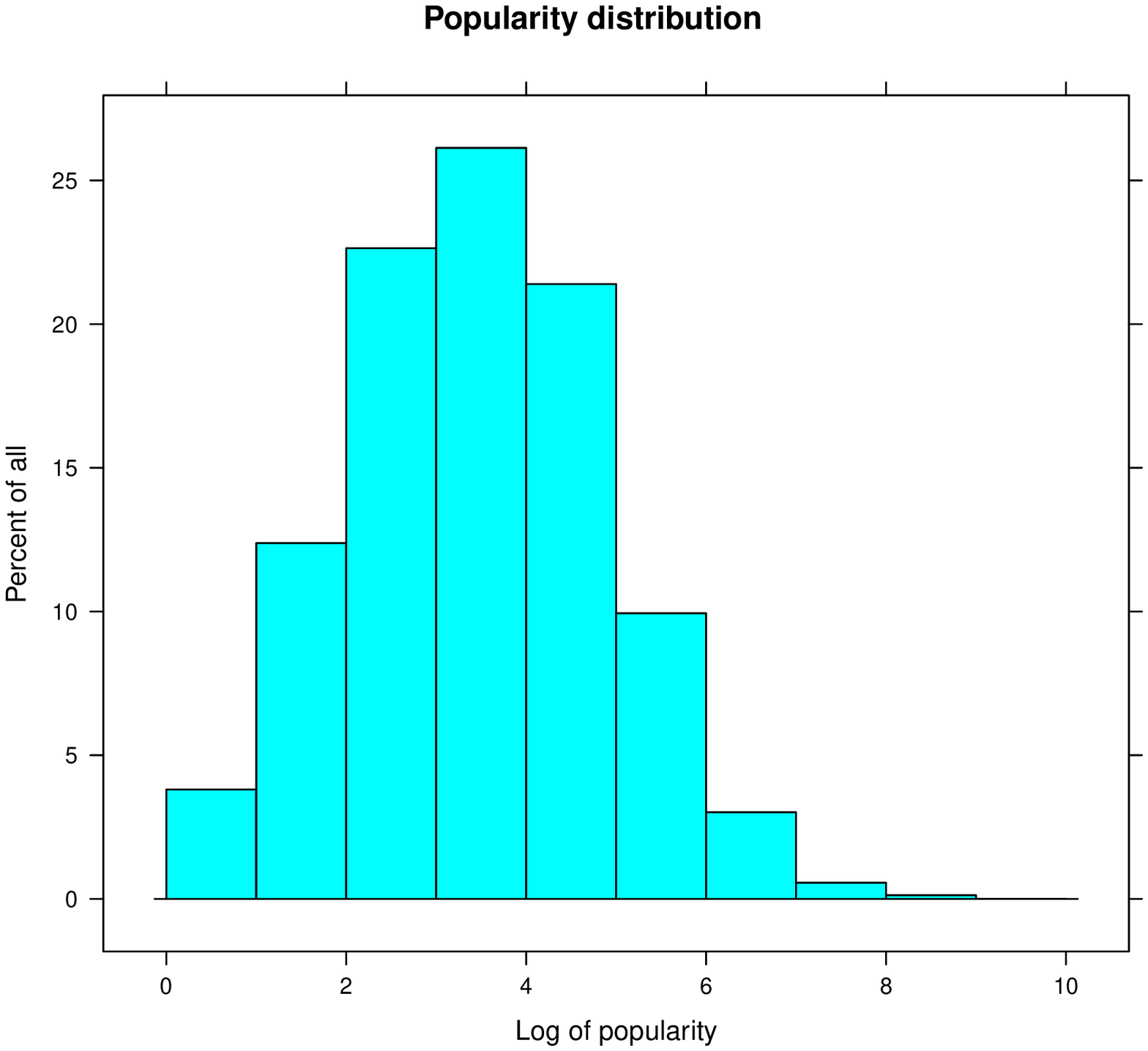}}
  \caption{Some features distributions from the YouTube dataset.}
  \label{features_dist}
\end{figure*}

As described in section{~\ref{data_how}}, we collected meta-data of more than $80000$ videos in a {MongoDB} database. In addition of the dynamics of
view-count used for modelling, the features we consider for each video are: the age (in number of days), the YouTube category and 
the popularity (i.e the total number of views at the day of crawling). Table{~\ref{cat_list}} lists the YouTube categories contained within the dataset
and Figure{~\ref{cat_dist}} shows their distribution. Table{~\ref{data_summary}} summarises the values of age and popularity metrics. Figure{~\ref{age_dist}} gives the ages
distribution and Figure{~\ref{logPop_dist}} is about the distribution of popularity (in $log$ values).

\begin{table}[h!]
  \centering
  \caption{List of all YouTube categories found inside the dataset}
  \vspace{.1cm}
  \begin{tabular}{|l|l|l|} \hline
    1. "Animals"	& 7. "Games" 	  & 13. "Shows"	  \\
    2. "Autos" 		& 8. "Howto"	  & 14. "Sports"  \\
    3. "Comedy"	  	& 9.  "Music"  	  & 15. "Tech"    \\
    4. "Education"     	& 10. "News" 	  & 16. "Travel"  \\
    5. "Entertainment"  & 11. "Nonprofit" &		  \\
    6. "Film"   	& 12. "People" 	  &	 	  \\ \hline
   \end{tabular}	
   \label{cat_list}
\end{table}


\begin{table}[h!]
 \centering
 \caption{Summary of age and popularity values within the YouTube dataset} \label{data_summary}
  \vspace{.2cm}
 \begin{tabular}{|l|l|} \hline
  \multicolumn{1}{|c|}{\textbf{Age (in days)}} & \multicolumn{1}{|c|}{\textbf{Popularity (number of views)}}\\ \hline
  Min: $5$		&Min: $1$\\ 
  $1$st Qu.: $140$	&$1$st Qu.: $2,650.10^{2}$ \\ 
  Median: $393$ 	&Median: $2,728.10^{3}$ \\ 
  Mean: $610,5$ 		&Mean: $6,091.10^{5}$ \\ 
  $3$rd Qu.: $923$	&$3$rd Qu.: $2,630.10^{4}$ \\ 
  Max: $2426$ 		&Max: $1,746.10^{9}$ \\ \hline
 \end{tabular}
\end{table}

%

\subsection{Data fitting}

\paragraph{Observations and normalisation}
For the data fitting, we only use the cumulative evolution of view-count. Given the evolution of view-count of a YouTube 
video as a function of time, 
we define a set of observations as : 
$(y_{i}, t_{i})_{1\leq i\leq n}$
where $y_{i}$ is the 
view-count at day $t_{i}$ and $n$ is the number of observations (this is also its age in number of days).
In order to avoid some technical issues due to the estimation algorithms, we use normalised observations : 

$(\frac{y_{i}}{y_{n}}, \frac{t_{i}}{t_{n}})_{1\leq i\leq n}$

\paragraph{Parameters estimation methods}\label{ParamEst}

We estimate the parameters of the models described in section~\ref{models} using regression algorithms both based on the mean squares criterion minimisation. 
Given a normalised set of 
observations $(y_{i}, t_{i})_{1\leq i\leq n}$, let $S$ be the expression for one model. The mean squares criterion ($MSC$) is then given by :
$$MSC = \sum_{i}(S(t_{i}) - y_{i})^2$$

We implemented two algorithms  
in order to automatically classify the dynamics of any video from YouTube in one of the models presented in the section~\ref{models}.

The first method is a simple linear regression. It works for videos where the view-count grows linearly over time $t$. 
In that case, the coefficient of determination $R$ gives a measure for the goodness of the fit : 
\[R = 1 - \frac{\sum_{i}(y_{i} - S(t_{i}))^2}{\sum_{i}(y_{i} - \bar{y})^2}\]
where $\bar{y}$ is the mean of $(y_{i})_{i}$.  In our experiments, we consider that a linear model is relevant if the value of $R$ satisfies  $\lvert R \rvert \geq 0.985$.
In the dataset, there are very few video dynamics that meet the linear case.  

The second method is the Levenberg-Marquardt Algorithm~\cite{LMA} which is known to be very efficient for the non-linear case. 
It is an iterative process for estimating parameters of the model through a minimisation problem of the $MSC$. 
Explicit formulation of the models have to be known to use this algorithm because the partial derivatives have to be calculated during 
the iterative process. One drawback of this method, like all other non-linear regression methods,  is that the solution could not be global but only a local one. 
Nevertheless, the Levenberg-Marquardt Algorithm suits very well for our models.

\begin{figure*}[t]
\centering
\subfloat[YouTube content with a concave shape \label{yt_negExp}]
{\includegraphics[width=0.17\textwidth]{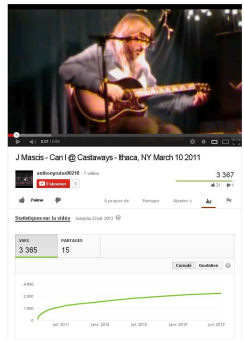}}\quad
\subfloat[Viewcount curve (plain) is approximated by a negative exponential curve (dashed) \label{negExp_fitting}]
{\includegraphics[width=0.26\textwidth]{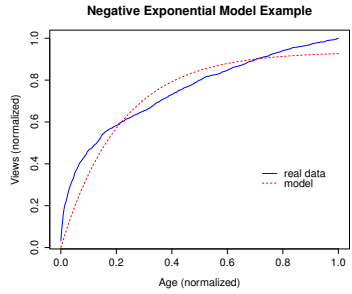}}\quad
\subfloat[Viewcount curve (plain) is approximated by a modified negative exponential curve (dashed) \label{ModNegExp_fitting}]
{\includegraphics[width=0.26\textwidth]{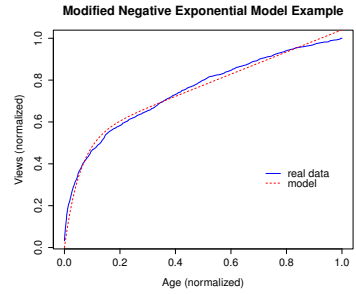}}
\caption{From a YouTube content (on the left side), parameters of the negative exponential model are estimated, then the obtained curve is compared to data (in the centre). 
The same process is applied to a negative exponential model in which a linear component has been added (on the right side)}
\label{NegExp}
\end{figure*}

\paragraph{Data fitting for Non viral contents}
A non viral behaviour is modelled by the dynamic equation~{\ref{negExp1}}. 
This model, called the negative exponential model, fits for contents which view-count curve goes concave then reaches an asymptote. 
In Figure~\ref{NegExp} we show an example where this model is applied. We observe
that the estimated curve (dashed line) admits an asymptote (which limit value is in fact the $M$ parameter of the model). 
This is typically  the case with  fixed finite population. But at the horizon, the curve that represents data (plain line)
seems to follow a line with a non zero slope. This is what we call the immigration phenomena, when the size of target population increases linearly in time. In that case, we
model the dynamics by the modified negative exponential functions introduced in subsection~\ref{nv_imm}:
\[S(t) = S(0) +  (M - S(0))(1 - e^{-\lambda t}) + kt\]
This model fits better as it is shown in Figure~\ref{ModNegExp_fitting}.
A mixed strategy, which consists in cutting data into two subsets and then applying linear model on one subset and negative exponential model on the other, will be discussed
later.

\begin{figure*}
 \centering
  \subfloat[YouTube video \label{psy_yt}]{
  \includegraphics[width=0.17\textwidth]{figeps/PsyGNS.eps}}
  \subfloat[Logistic model fitting \label{sigmoid_fitting}]{
   \includegraphics[width=0.26\textwidth]{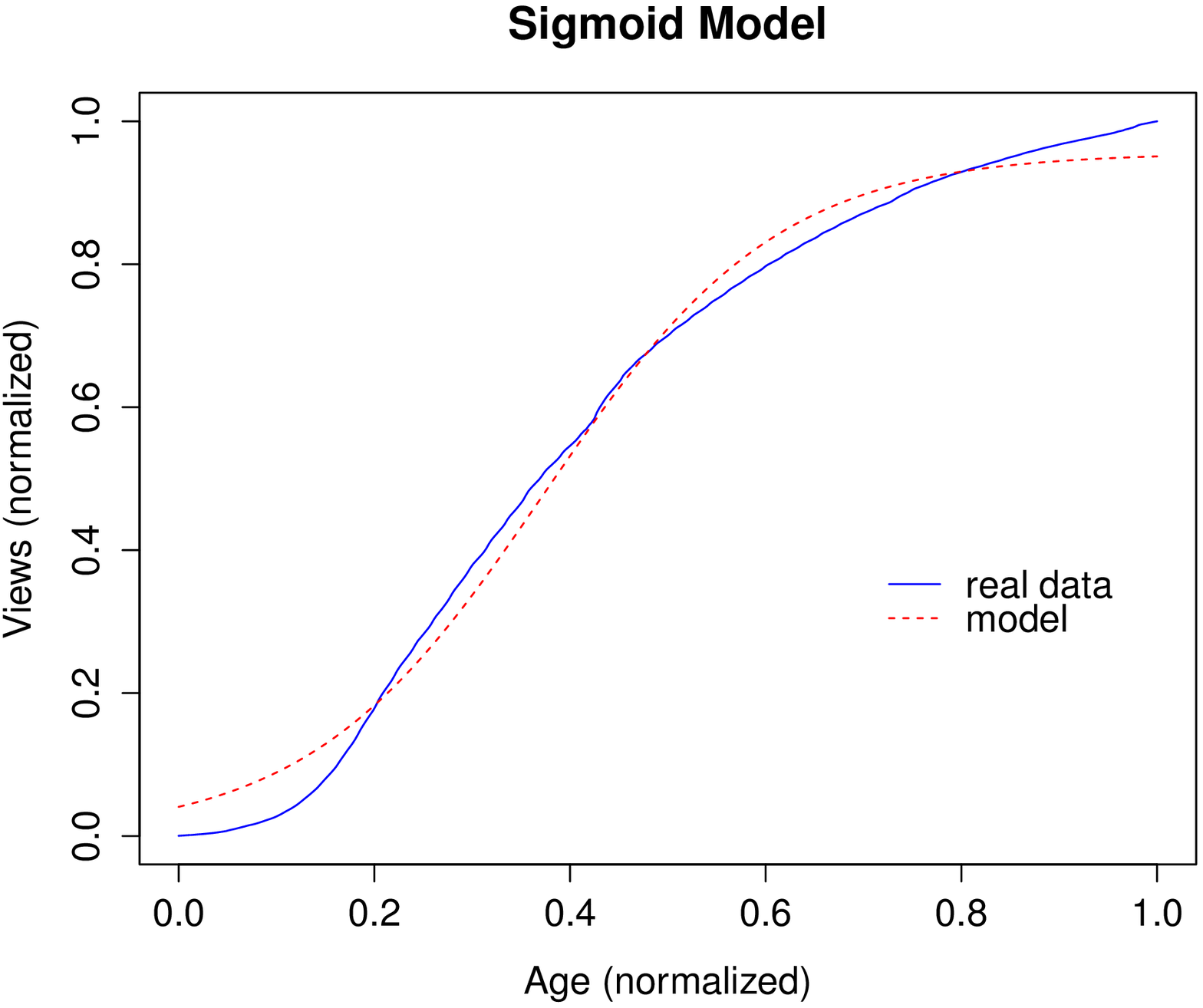}}
  \subfloat[Gompertz model fitting \label{gompPsy}]{
  \includegraphics[width=0.26\textwidth]{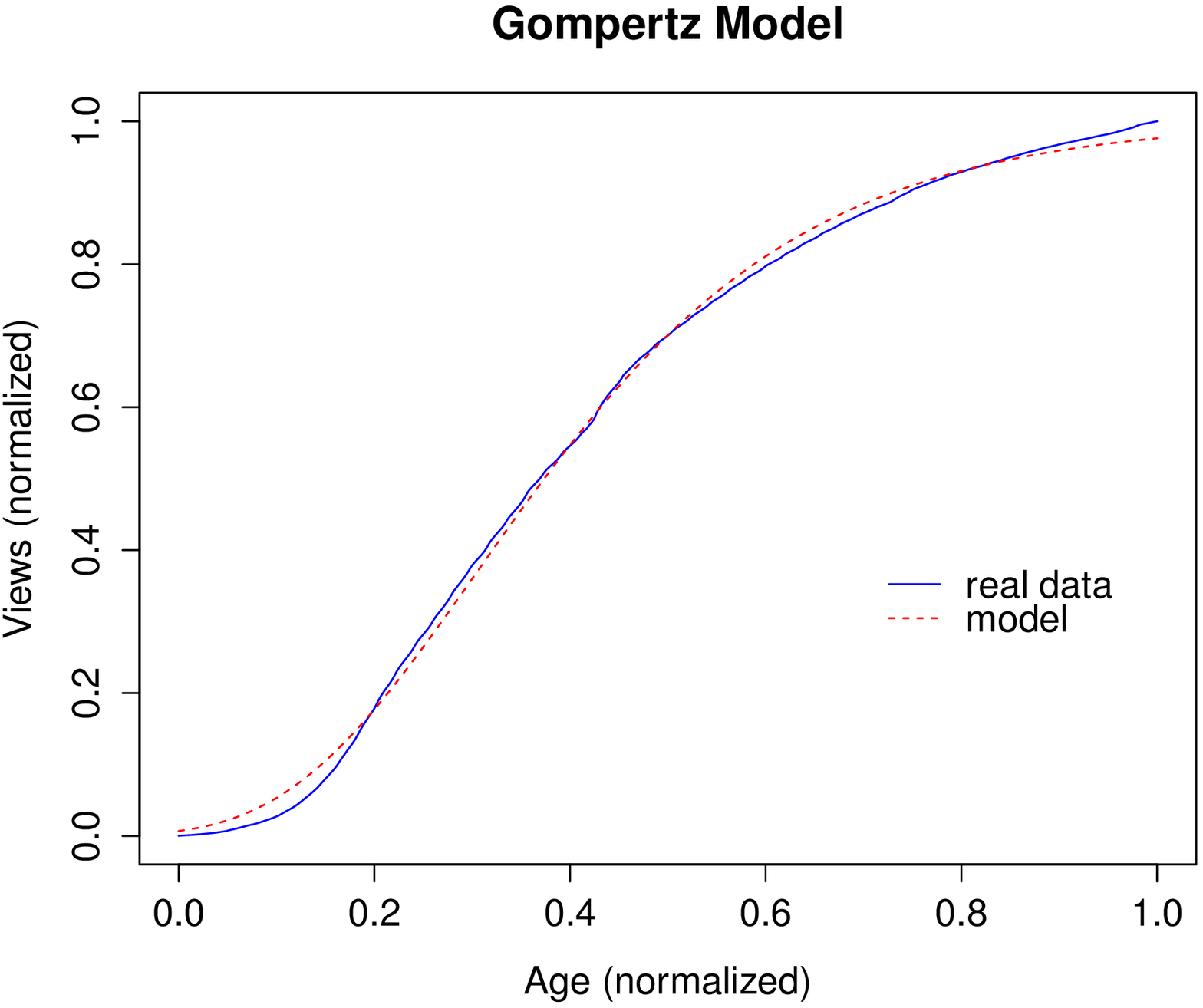}}
  \subfloat[Modified Gompertz model fitting \label{GompLinPsy}]{
  \includegraphics[width=0.26\textwidth]{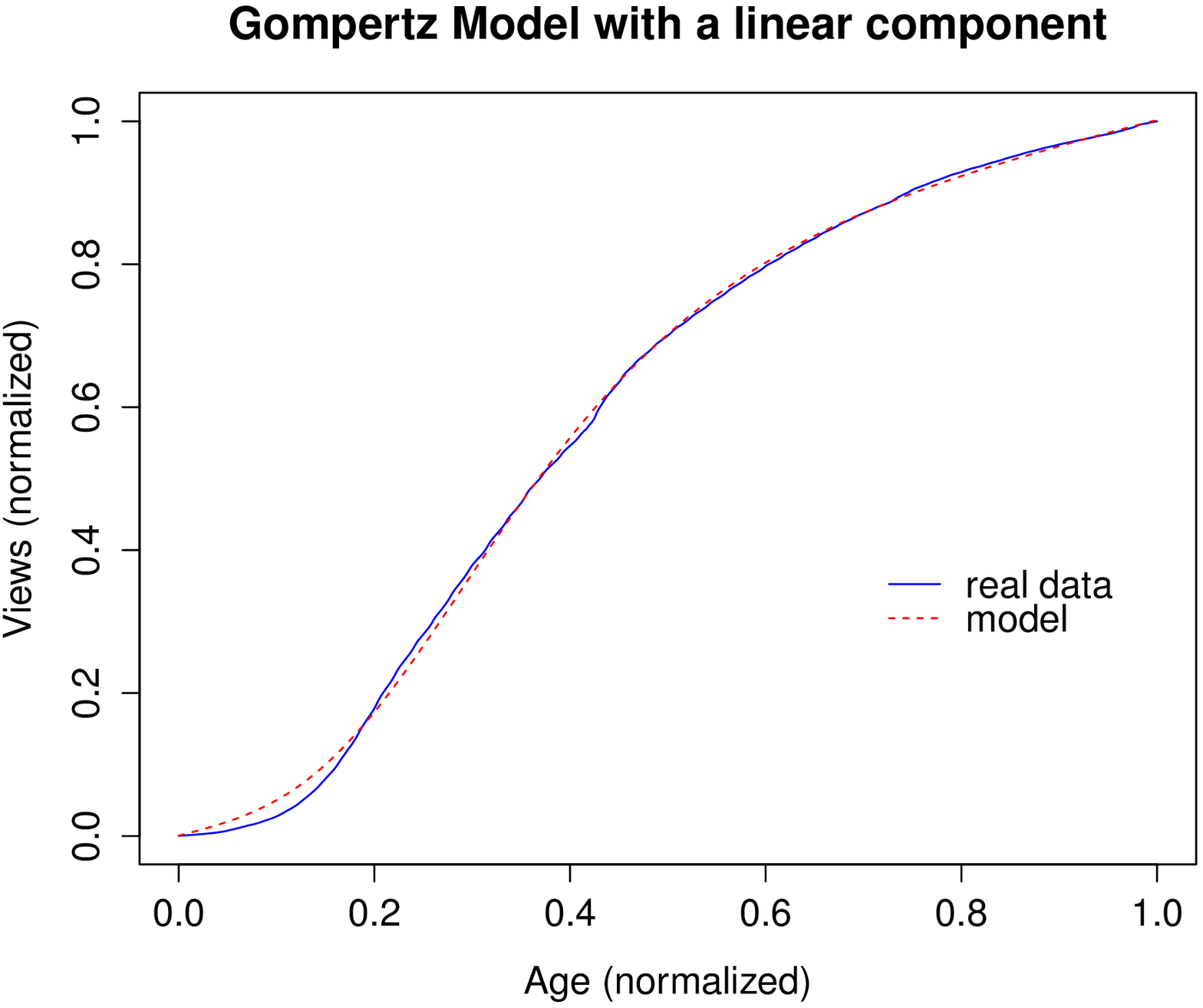}}
  \caption{From a YouTube video with a S-shaped view-count curve (~\ref{psy_yt}), we first fit the logistic model in~\ref{sigmoid_fitting}. The estimated curve (dashed) is
compared with the actual normalised view-count curve (plain). Then the same is done with the Gompertz model in~\ref{gompPsy} and finally with the 
modified Gompertz model in~\ref{GompLinPsy}.}
  \label{viral_fits}
\end{figure*}

\paragraph{Data fitting for Viral contents}
Three models have been considered in the case of viral contents: logistic model, Gompertz model and a modified Gompertz model (see section~\ref{vir_imm}). The first two 
models are for the context of fixed finite population and the third one is introduced in the case of immigration.\\
Figure~\ref{viral_fits} is an example where we fit these models to one YouTube content (Figure~\ref{psy_yt}). We observe that the S-shape of the logistic model curve 
is symmetric due to the symmetrical property of sigmoid function (Figure~\ref{sigmoid_fitting}).
However, the convex phase and the concave phase are non symmetric as we can observe in Figure~\ref{psy_yt}. Hence the Logistic model does not fit well.
Then, Gompertz model and modified Gompertz model are fitted to the same YouTube content. 
The Gompertz model (Figure~\ref{gompPsy}) fits better than the logistic model, and the modified Gompertz model (Figure~\ref{GompLinPsy}) 
describes better the behaviour of the data at the horizon (immigration phenomena).



\paragraph{Mixing linear and non-linear models}
An issue that results from the model we use is the changes of the curve dynamics at the horizon.
We observe two types of behaviour at the horizon : a flat line showing that the limit of the potential population has been reached or an oblique line highlighting the 
fact that the population continues to grow.

\begin{figure}[ht!]
\centering
\subfloat[YouTube video (Obama about gay marriage) \label{Ob_yt}]{
\includegraphics[width=0.17\textwidth]{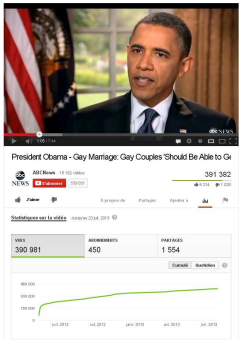}}
\subfloat[Mixed linear/non linear fitting \label{Mixed_fitting}]{
\includegraphics[width=0.6\columnwidth]{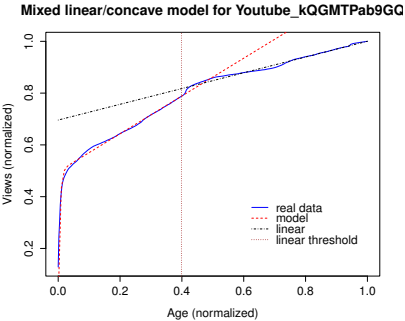}}
\caption{From a YouTube content (Figure~\ref{Ob_yt}), a mixed procedure is made to estimate a linear model and a non linear model on two subset of data
(Figure~\ref{Mixed_fitting})}
\label{MixedModel}
\end{figure}

The first case is coherent with the description of dynamics (then the maximum is one of the parameters of 
the dynamic equation of the model). However, for the second case we need to add a linear component to the solution function, implying that the initial dynamic equations 
 are not valid anymore.
\\
Hence, we propose to adopt a two phases approach for 
data fitting. Figure~\ref{MixedModel} gives an example where this method is 
applied for a YouTube content (Figure~\ref{Ob_yt}).

Given a set of observations $(y_{i}, t_{i})_{1\leq i\leq n}$, a first phase consists in pointing out a linear behaviour from a 
time $t=t_{k}, k \in {1,..,n}$. The idea is to find $k$ in an iterative way in order to have a good regression line for the subset $(y_{i}, t_{i})_{k\leq i\leq n}$.
The algorithm is the following : we first fix a threshold $\epsilon$ reasonably small.
We then apply a linear regression with all the points. The regression line gives the set $(\hat{y}_{i})_{i}$ of the estimated values. Let $\bar{y}$ be the mean of $(y_{i})_{i}$,
we consider the coefficient of determination $r_{1}$ given by:
\[r_{1} = 1 - \frac{\sum_{i=1}^{n}(y_{i} - \hat{y}_{i})^2}{\sum_{i}(y_{i} - \bar{y})^2}\]
\begin{itemize}
 \item if $\lvert1 - r_{1}\rvert\leq \epsilon$,  observations could be considered as linear (the whole view-count curve could be well described by the linear model) and the process ends.
 \item else, the first element of $(y_{i}, t_{i})_{1\leq i\leq n}$ is removed from the set and a new linear regression is done for the subset $(y_{i}, t_{i})_{2\leq i\leq n}$. That gives
a new coefficient of determination $r_{2}$
\end{itemize}
The process is repeated until the coefficient of determination $r_{k}$ satisfies $\lvert1 - r_{k}\rvert\leq \epsilon$. Doing that, we determine the rank $k$ from 
which observations $(y_{i}, t_{i})_{k\leq i\leq n}$ can be considered as having a linear behaviour. In Figure~\ref{Mixed_fitting}, time $t_{k}$ is represented by a 
vertical line. From $t_{k}$ the behaviour can be well described by a linear model (dot-dashed line).
\\
In a second phase, parameter estimation for the non-linear models presented in previous section is done on the subset $(y_{i}, t_{i})_{1\leq i\leq k}$. 
Figure~\ref{Mixed_fitting} illustrates this phase : here, the modified negative exponential model is applied to the subset on the left side of $t_{k}$ (dashed line).



\section{Automatic classification}

In this section we first investigate the process for an automatic classification of YouTube contents. We then go further in analysing results of our experiment.
Finally we give some keys of how to use this classification in order to predict the view-count.

\subsection{Classification issues}
The main goal of our work is to provide a system that can automatically classify YouTube contents by associating one model to one content. For each content, two issues 
have to be managed : 
First, evaluate each model in order to know which models are good candidates. 
Then compare candidates to determine which one is the best.

Let us consider first the question of evaluating each model. As explained in section{~\ref{ParamEst}}, we perform parameters estimation based on the least squares 
criterion minimisation. 
Define the mean error rate ($MER$):
\[MER = \frac{1}{n}\sum_{i}\frac{\lvert S(t_{i}) - y_{i}\rvert}{y_{i} + 1}\]
$MER$ criterion is the mean error rate done by the model regarding the observations. For example, if $MER\leq0,05$, it can be said that on average, the estimation 
error is under $5$\% of the observed value. With this criterion we can fix a threshold beyond which one model would be considered as unreliable.
Actually, the expected mean error rate should be $\frac{1}{n}\sum_{i}\frac{\lvert S(t_{i}) - y_{i}\rvert}{y_{i}}$. $MER$ is exactly this calculation applied to 
estimated data and observed data which have been translated by one on the y axis. Since the observed time series are normalised and so bounded between $0$ and $1$,
$MER$ definition avoids generation of large errors due to some very small values of $y_i$ and also avoids division by zero when $y_i$ is equal to $0$.
\\
In order to compare models for which $MER$ is lesser than a certain threshold, we introduce a criterion of quality discussed in \cite{ChiTest}. To formulate this 
criterion we first define the degree of freedom of a model ($df$). Let $p$ be the number of parameters of the model, the degree of freedom is then defined by:
$df = n-p$.
The criterion of quality, named ``goodness of fit'' ($GoF$) is then given by:
$$GoF = \frac{1}{(df)}MSC$$
The model which has the smallest $GoF$ will be considered as the best one. In other words, it will be the model that best fits the data.
\begin{table}
 \centering
 \caption{Goodness of fit for models from Figure~\ref{NegExp}} \label{TableNegExp}
\vspace{.2cm}
 \begin{tabular}{|c|c|c|c|} \hline
  \textbf{Model} & \textbf{$MCS$} & \textbf{$MER$} & \textbf{$GoF$}\\ \hline
  Neg. exponential & $3,558$ & $0,074$ & $0,004$\\ \hline
  Modified neg. exponential & $0.453$ & $0,027$ & $4,98.10^{-4}$\\ \hline
 \end{tabular}
\end{table}
In Table~\ref{TableNegExp} and Table~\ref{TableGomp} we give values of $MSC$, $MER$ and $GoF$ for models used respectively in Figure~\ref{NegExp} and 
Figure~\ref{viral_fits}.
In the example from Figure~\ref{NegExp}, with a threshold of $MER$ fixed at $0,075$, both negative exponential model and modified negative exponential model are relevant.
With a $GoF$ of value $4,98.10^{-4}$, modified negative exponential model is the one that best fits the data. In the case of YouTube content depicted in 
Figure~\ref{viral_fits}, if the threshold of $MER$ is fixed at the value of $0,02$, sigmoid model is not reliable whereas Gompertz model and modified Gompertz model are 
under the threshold. According to the value of $GoF$, modified Gompertz model is the best with $GoF = 8,831.10^{-5}$. Further, the issue of fixing a value for the $MER$ 
threshold is crucial to rely on an acceptable filter for several videos.
\begin{table}
 \centering
 \caption{Goodness of fit for models from Figure~\ref{viral_fits}} \label{TableGomp}
 \vspace{.2cm}
 \begin{tabular}{|c|c|c|c|} \hline
  \textbf{Model} & \textbf{$MCS$} & \textbf{$MER$} & \textbf{$GoF$}\\ \hline
  Sigmoid & $0,480$ & $0.021$ & $10^{-3}$\\ \hline
  Gompertz & $0,092$ & $0,018$ & $1,846.10^{-4}$\\ \hline
  Modified Gompertz & $0,033$ & $0,008$ & $8,831.10^{-5}$\\ \hline
 \end{tabular}
\end{table}


\subsection{Results analysis}
\begin{figure}[ht!]
\centering
\subfloat[Percent of contents by bins of $MER$ values\label{MEROverall}]{\includegraphics[width=0.5\columnwidth]{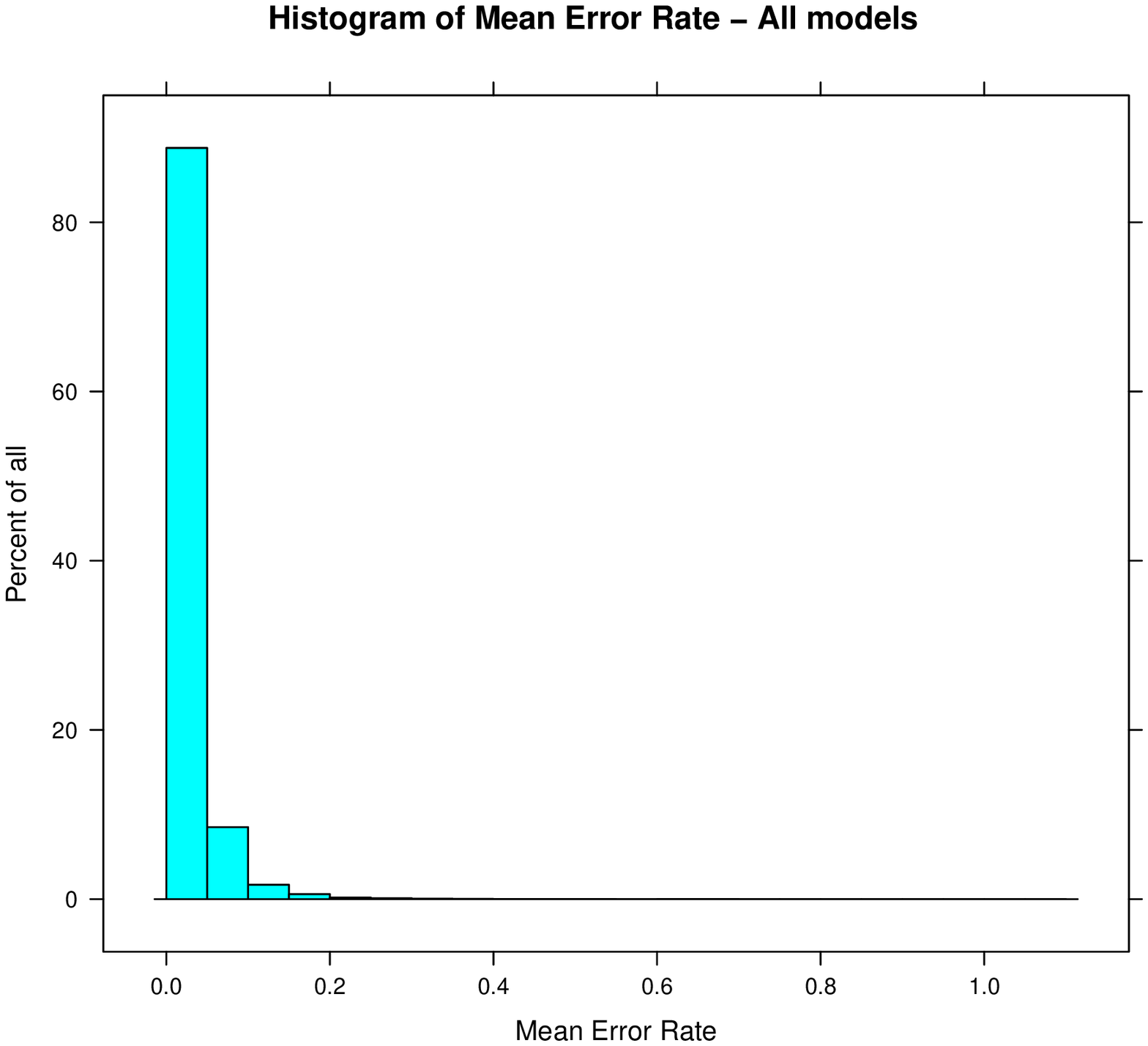}}
\subfloat[Models distribution after classification over the whole dataset\label{modDist}]{\includegraphics[width=0.5\columnwidth]{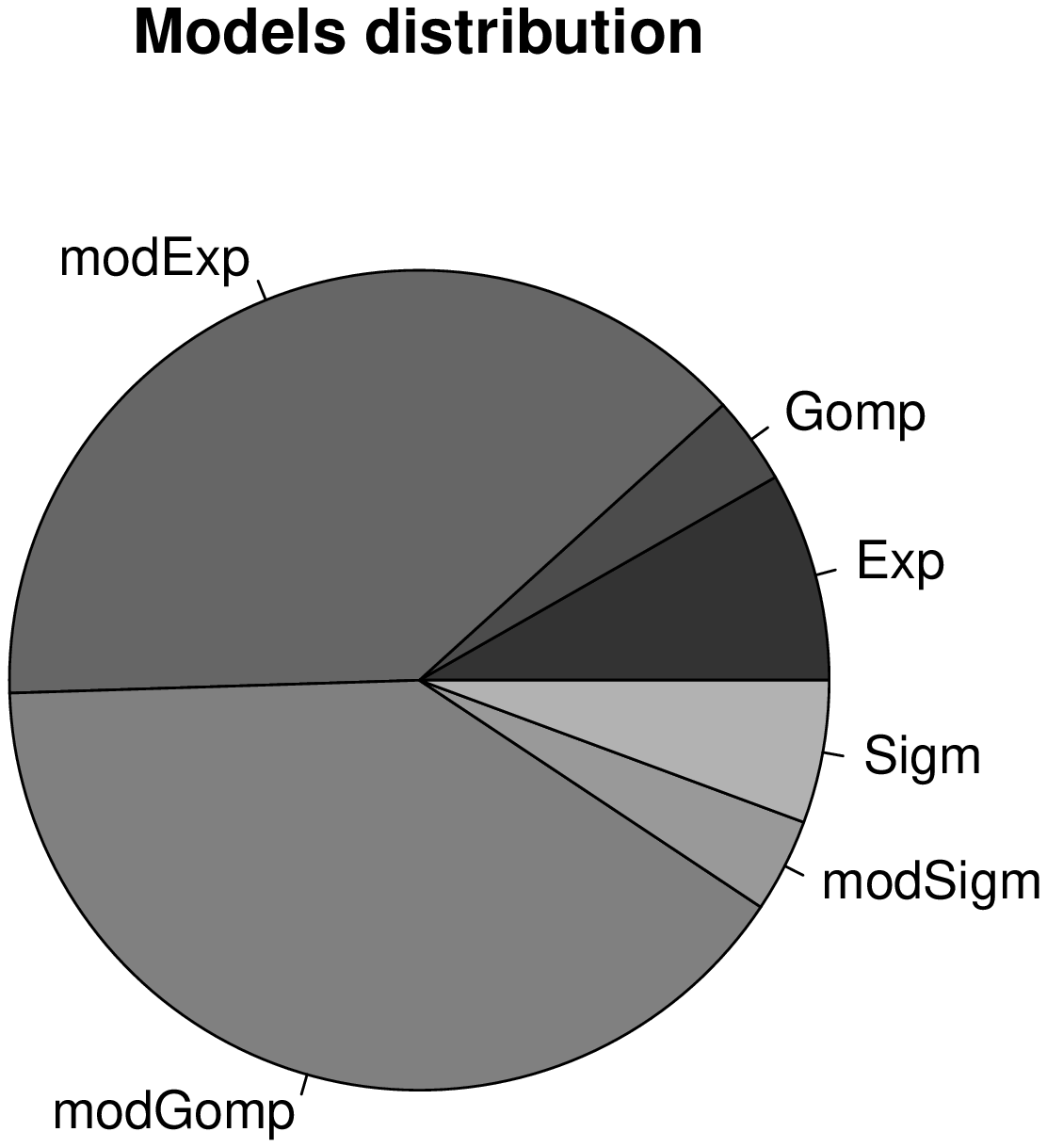}}
\vskip0cm
\subfloat[$MER$ distribution for each model\label{MER_by_mod}]{\includegraphics[width=0.5\columnwidth]{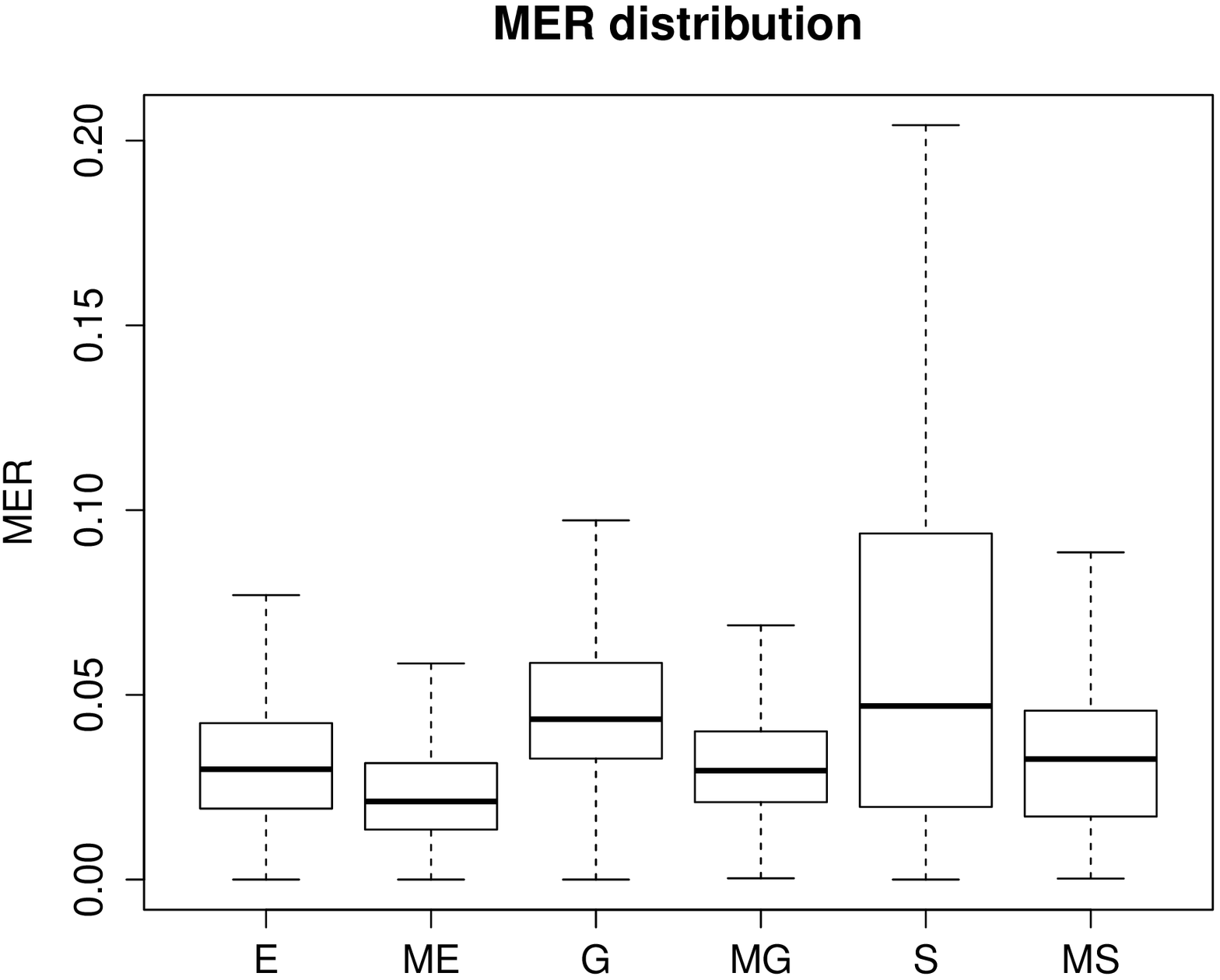}}
\caption{Sample analysis of an automatic classification for dissemination processes in YouTube}
\end{figure}

Given one video from the dataset, after doing parameters estimation for each model, the system automatically associates one model to the video using $MER$ and $GoF$ 
criteria. 
We give results about the distribution of $MER$ values in Figure{~\ref{MEROverall}}. 
It is shown that $89\%$ of videos are associated to a model with a $MER$ under $0,05$.
A mean error of $5\%$ seams reasonable to consider a reliable fitting. Hence, if $MER$ is fixed at $0.05$, we should conclude that the classification system is 
efficient in almost $90\%$ of the cases.

\begin{figure}[ht!]
\centering
\subfloat{
\includegraphics[width=0.17\textwidth]{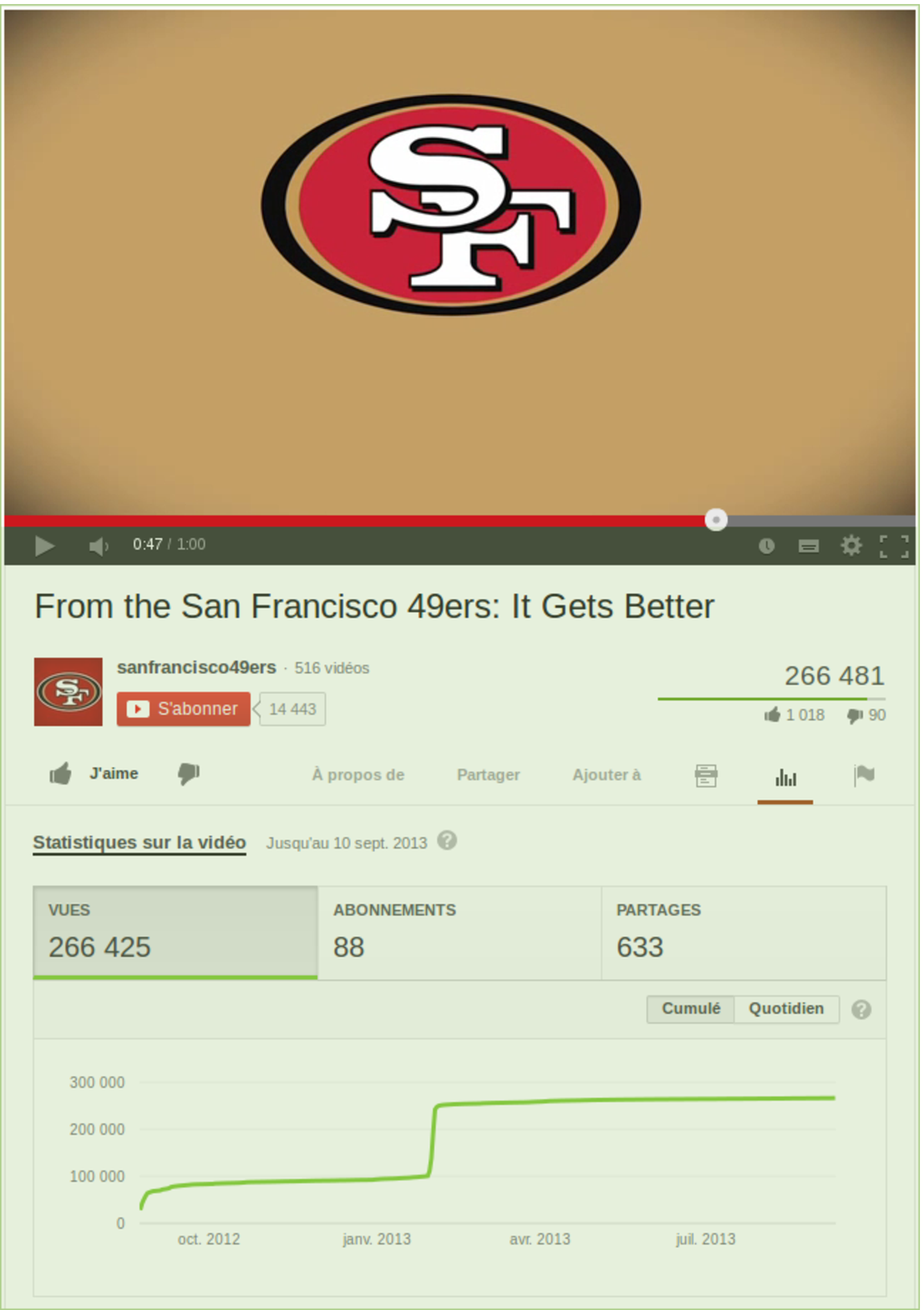}}
\subfloat{
\includegraphics[width=0.6\columnwidth]{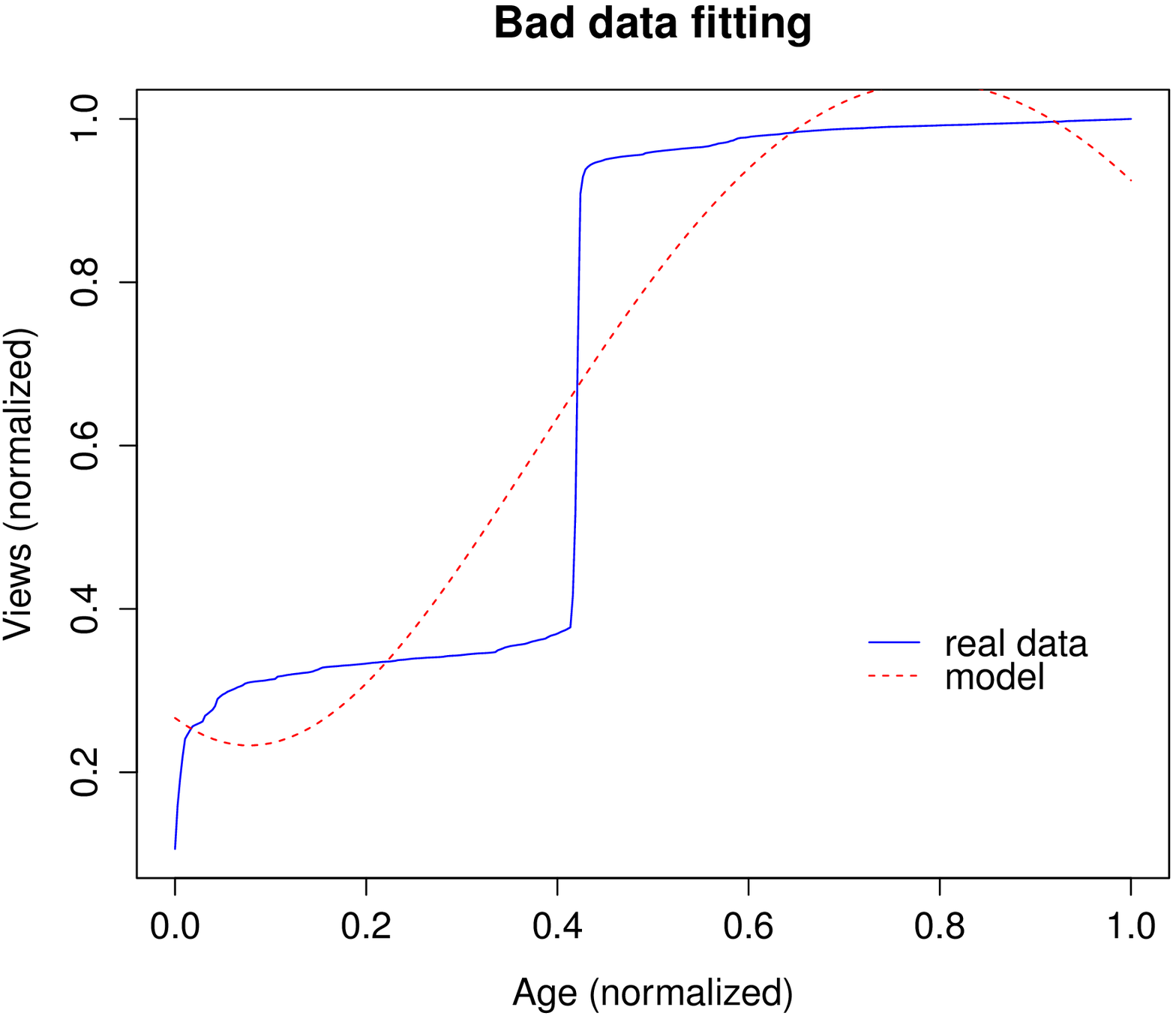}}
\caption{One bad fitting example.}
\label{badVideo}
\end{figure}

Note that if the threshold of $MER$ is fixed at $0,1$, 
more than $97\%$ of the videos correspond to one of the models. There is at most $2\%$ of the videos for which the association to one 
of our models gives a high error rate (let say more than $10\%$). Figure~\ref{badVideo} illustrates one example of such a video.\\
In this case, the system associates the modified Gompertz model to the video, with a $MER$ equal to $0,127$ and a $GoF$ of value $1,13.10^{-1}$. 
The association is absolutely unreliable due to the many sharp changes of the behaviour. Indeed, it seems that the models are unable to capture
the effect of multiple peaks in view-count evolution. This might be the case of view-count curves that are somehow 'non differentiable'. Further investigation
has to be made about this issue, this is one of our direction in future works. \\

In Figure{~\ref{MER_by_mod}} we focus on the $MER$ distribution for each model. Note first that we did not used the linear model because it insignificantly 
appears in the dataset. Secondly we introduce a new model, named modified sigmoid model, given by the logistic model with a linear component added 
(as done in section{~\ref{vir_imm}} with the modified Gompertz model).


It appears that modified negative exponential and modified Gompertz models give fitting with less error than the other in most of the cases: they 
can be referred as the most reliable models for our dataset. 
We present 
the models distribution in Figure{~\ref{modDist}}.
The same two models: modified negative exponential and modified Gompertz, almost cover the whole dataset with the same amount of videos for each one. 
Regarding their reliability, it enforces the classification efficiency.
The former is a non viral model whereas the latter is viral, meaning that there is quite a balance between viral and non viral contents. 
Both highlight an immigration process, leading to the conclusion that a lot of YouTube contents still attracting viewers even after a long period.

\begin{table*}[bht]
 \centering
 \caption{Confidence Intervals for models distribution proportion} \label{TableCI}
 \vspace{.2cm}
 \begin{tabular}{|c|c|c|c|} \hline
  \textbf{Model/sampling} & \textbf{$1000$} & \textbf{$10000$} & \textbf{$Whole dataset (81657)$}\\ \hline
  Exponential & $(0.05730450 - 0.07221 - 0.09049091)$ & $(0.06681688 - 0.071761 - 0.07703725)$ & $(0.07009212 - 0.07184932 - 0.07364694)$\\
  Modified Exponential & $(0.3390106 - 0.36885 - 0.39971)$ & $(0.3584859 - 0.367935 - 0.3774861)$ & $(0.3648174 - 0.3681252 - 0.3714455)$\\
  Sigmoid & $(0.02141720 - 0.03089 - 0.04411793)$ & $(0.02735499 - 0.030602 - 0.03421483)$ & $(0.02928192 - 0.03044442 - 0.03165132)$\\ 
  Modified Sigmoid & $(0.02221527 - 0.03185 - 0.04522983)$ & $(0.02779605 - 0.031068 - 0.03470537)$ & $(0.02982271 - 0.03099551 -0.03221266)$\\ 
  Gompertz & $(0.01530999 - 0.02342 - 0.03536814)$ & $(0.02065726 - 0.023495 - 0.02670484)$ & $(0.02260972 - 0.02363545 - 0.02470626)$\\
  Modified Gompertz & $(0.3310334 - 0.3607 - 0.3914506)$ & $(0.3535128 - 0.362832 - 0.3723571)$ & $(0.3595007 - 0.362798 - 0.3661083)$\\ 
  \hline
 \end{tabular}
\end{table*}

In order to assess the evidence provided by our dataset  on models distribution, we provide in Table~\ref{TableCI}  95\% confidence interval  for different  sample 
sizes involved in the study.   This table indicates that  the whole dataset  leads to very good precision on   models distribution.  Furthermore, a sampling with 10000  videos still gives an  accurate estimate of this proportion. 

Now it is natural to ask whether the distribution of our classification is still the same with respect to  main categories  in YouTube   and popularity of a video.

\begin{figure}
\centering
\includegraphics[width=0.26\textwidth]{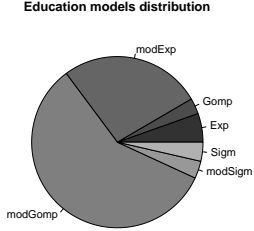}
\caption{Models distribution for the Education category}
\label{EducationMD}
\end{figure}

In the case of categories, we consider the different YouTube categories into the models classification: 
Four main categories can be highlighted regarding the distribution in Figure{~\ref{cat_dist}}.
Music (over $14000$ videos), Entertainment (over $8500$ videos), People (around $7500$ 
videos) and Education (almost $6000$ videos). 
In general, the models distribution in each category is not far from the distribution considering all categories combined. However, Education category is quite different. 
The models distribution
for this category is given in Figure{~\ref{EducationMD}.}
%
%
We note that the modified Gompertz model dominates. Furthermore, viral models cover almost $75\%$ of the videos.
One might assume that Education is a word-of-mouth category where videos dissemination results mainly from viewers 
influence and few benefits from advertising processes or internal YouTube mechanisms such as recommendation system.

We now analyse the models distribution
considering the popularity of videos. Let us introduce the different classes of popularity defined for our dataset:
According to the distribution depicted in Figure~\ref{logPop_dist}, we define seven classes of popularity listed in Table~\ref{pop_classes}.
\begin{table}[ht!]
 \centering
 \caption{Popularity classes} \label{pop_classes}
 \begin{tabular}{|c|c|} \hline
  \textbf{Popularity class}	& 	\textbf{Total number of views $V$} \\ \hline
  Extremely unpopular (EUP)		&	$0 \leqslant V < 10$ 		\\ \hline
  Very unpopular (VUP) 		& 	$10 \leqslant V < 100$ 		\\ \hline
  Unpopular (UP) 			& 	$100 \leqslant V < 1000$ 	\\ \hline
  Not so popular (NSP) 		& 	$1000 \leqslant V < 10^{4}$ 	\\ \hline
  Popular (P)			&	$10^{4} \leqslant V < 10^{5}$ 	\\ \hline
  Very popular (VP) 			&	$10^{5} \leqslant V < 10^{6}$	\\ \hline
  Extremely popular (EP) 		& 	$10^{6} \leqslant V$		\\ \hline
 \end{tabular}
\end{table}
We show the models distribution for each popularity class in Table~\ref{modDist_pop_table}.\\
\begin{table}[ht!]
 \centering
 \caption{Models distribution by popularity class (in \%)} \label{modDist_pop_table}
 \begin{tabular}{|c|c|c|c|c|c|c|c|} \hline
  \textbf{Model}   & \textbf{EUP}   & \textbf{VUP}  & \textbf{UP}  & \textbf{NSP} & \textbf{P}	& \textbf{VP}	& \textbf{EP}   \\ \hline
  Exp		   &    $11.4$ 	    &$12.2$ 	    &  $8.4$  	   &  $8$  	  & $6.8$	& $6.2$		& $5.7$  	\\ \hline
  Gomp	   	   &	$1.6$       &$2.8$  	    &  $1.8$  	   &  $2.5$  	  & $3.6$	& $3.2$		& $2.1$  	\\ \hline 
  ModExp	   &	$11.6$      &$54.5$ 	    &  $48.9$ 	   &  $35.2$ 	  & $35.1$	& $42.3$	& $47.5$ 	\\ \hline
  ModGomp	   &	$2.5$       &$19.4$ 	    &  $34.7$ 	   &  $48.7$ 	  & $49.3$	& $44.3$	& $42.8$ 	\\ \hline
  ModSigm	   &	$1.8$       &$4.3$  	    &  $4.3$  	   &  $3.6$  	  & $2.9$	& $2.3$		& $0.8$  	\\ \hline
  Sigm	   	   &	$70.7$      &$6.6$  	    &  $1.5$  	   &  $1.7$ 	  & $2$		& $1.3$		& $0.8$  	\\ \hline
 \end{tabular}
\end{table}
We can observe that the distribution varies according to the classes of popularity. First of all, the logistic model (referred as sigmoid in the plots) dominates the
extremely unpopular category (constituted by videos of less than 10 views). These results are not reliable due to the few different values of the view-count for these videos. 
Actually, in our system, all the models should be good for those videos and the first that is tested is
the logistic model so that it is the one which is chosen. 
Popular class and not so popular class can be grouped in terms
of models distribution with around 50\% for modified Gompertz model and 35\% for modified exponential model. We can also group very popular class and extremely popular 
class for which distribution is slightly equivalent to the whole dataset distribution (see Figure~\ref{modDist}). Very unpopular and unpopular classes exhibit 
modified exponential model in around 50\% of the cases. The modified Gompertz model represents less than 20\% in very unpopular class whereas it covers almost 35\%
of the videos in unpopular class.

\subsection*{Classification models for  prediction}
In this section, we illustrate a mechanism  for predicting the future  evolution of view-count of a video.  In particular, we propose a simple model  that predicts 
the evolution of view-count from a given time  $t_f$ till a target date $t_p$ with $t_p>t_f$.  We call a prediction window $T$ the difference between $t_p$ and $t_f$. 
This prediction is based on  the early  historical information  of a video  which is  given by a set of   observations   $(y_i, t_i)_{1\leq i\leq f}$ till time $t_f$ where $f$ is number of obesrvations\footnote{The datasets used for prediction contains videos with at least $50$ days old.}.  Combining these information with  our classification models, the evolution of view-count is estimated using data fitting  in order to select a mathematical model.  Using our datasets, we evaluate  the maximum size of  prediction window  with at most 5\% mean error, i.e 
$$
T_{max} = \max \{t_p- t_f |  \frac{1}{p-f}\sum_{i=f+1}^{p}{\frac{|S_{t_f}({t_{i}}) - y_{i}|}{(y_{i}+1)}}  \leq  0.05 \}
$$
where $S_{t_f}$ is the selected mathematical model.    
\begin{table}[ht!]
 \centering
 \caption{Mean and variance of prediction window size in the half life scenario} \label{halflife_table}
 \begin{tabular}{|c|c|c|c|} \hline
Model	& mean	& var 	& number of videos\\ \hline
E	& $0.5833692$	& $0.1504571$	& $4132$\\ \hline
ME 	& $0.576914$	& $0.1415303$	& $25281$\\ \hline
G	& $0.3435265$	& $0.1160928$	& $683$\\ \hline
MG & $0.4596889$	& $0.1360676$	& $19349$\\ \hline
S & $0.6765688$	& $0.1544357$	& $1030$\\ \hline
MS & $0.4625144$	& $0.1280855$	& $1659$\\ \hline
 \end{tabular}
\end{table}
\begin{figure}[ht!]
 \centering
 \subfloat[50 days classification \label{delta50d_case}]{
 \includegraphics[width=0.5\columnwidth]{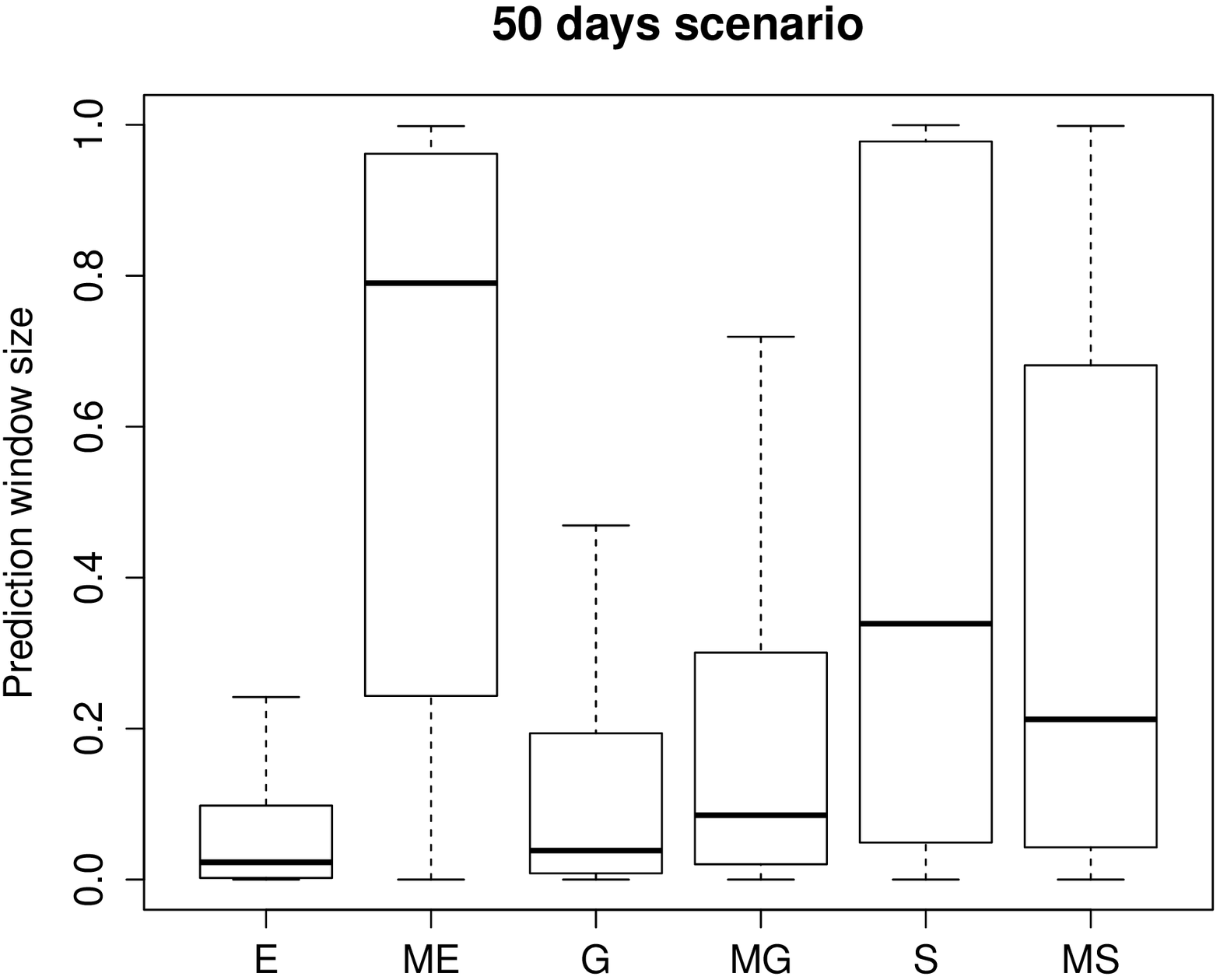}}
 \subfloat[Half life classification \label{deltaHalf_case}]{
 \includegraphics[width=0.5\columnwidth]{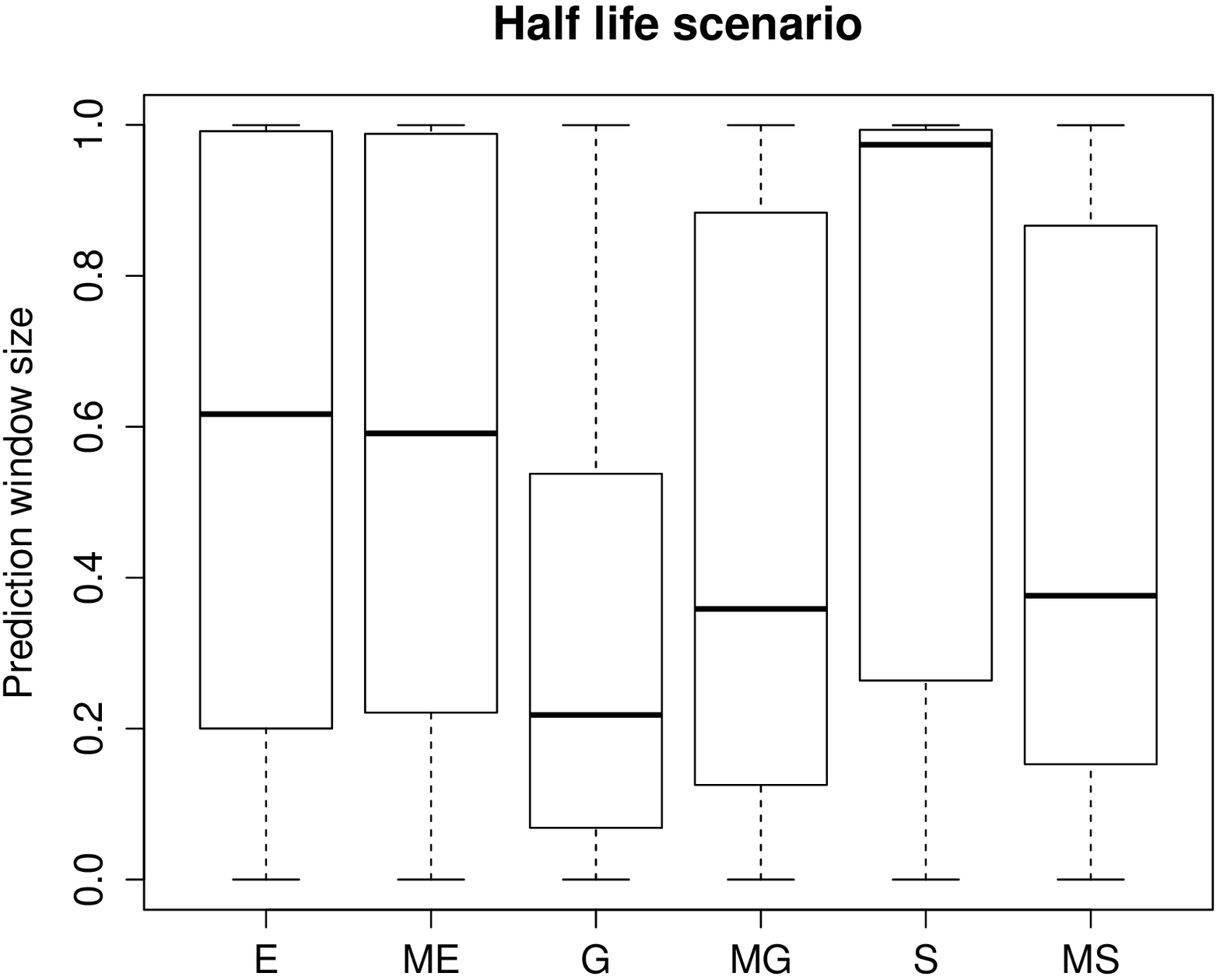}}
 \caption{Prediction window size according to models type}
 \label{window_size}
\end{figure}

We test our prediction for the scenario in which $t_f$ corresponds to   half life cycle.    Let $\Delta T=\frac{T_{max}}{t_n-t_f}$ where  $t_n-t_f$ is the remaining 
time of life cycle of a video from $t_f$.  Note that $\Delta T$ is bounded by $1$.  Fig.~\ref{deltaHalf_case}  depicts the mean and variance of  $\Delta T$  for each 
identified model.    Table~\ref{halflife_table} precises values of mean and variance for each model as well as the number of videos classified in the  different models.   
Our results  show that our prediction is very powerful  and most models provide a  prediction window that  long enough within an error bound at $5$\%.  Further  
we observe that  our scheme can perfectly  predict the evolution of view-count till the half of the remaining time of life cycle from the time of prediction.  

We tested here the prediction based on a learning sequence that was half the lifetime of each video in the dataset. This allows the prediction to rely on the same 
amount of data independently of the real duration of the video. We next compare this to the case in which, in contrast, the learning sequence has a fixed duration of 
50 days.  We note that $50$ days  represent much less than half the lifetime for most videos in the data set and therefore the prediction is less accurate. 
The corresponding  results of this scenario are  depicted in Table~\ref{50d_table}  and  Fig.~\ref{deltaHalf_case}. 
In spite of this problem we get  similar results of the average prediction window for models \textit{modified Gompertz} and \textit{sigmoid} (\textit{Logistic}). 


\begin{table}[ht!]
 \centering
 \caption{Mean and variance of prediction window size in the 50 days scenario} \label{50d_table}
 \begin{tabular}{|c|c|c|c|} \hline
Model	& mean	& var 	& number of videos\\ \hline
E	& $0.1240775$	& $0.06050271$	& $3401$\\ \hline
ME 	& $0.6159881$	& $0.1384898$	& $21788$\\ \hline
G	& $0.1861161$	& $0.09028605$	& $687$\\ \hline
MG & $0.2314134$	& $0.09438778$	& $13821$\\ \hline
S	& $0.4750911$	& $0.1083194$	& $1137$\\ \hline
MS & $0.2082774$	& $0.1798454$	& $1561$\\ \hline
 \end{tabular}
\end{table}



It's now natural to further investigate our prediction method in particular in case of early predictions. We slightly modify our method 
of calculating the prediction window size and consider values of $T$ equal to $7$ days, $15$ days and $30$ days. In each scenario,  we fix an horizon at 3 times 
the observed window size. 
We implement two ways for computing the prediction window size. The first one is the same as in the previous method and is called the soft window. The other one is called
the hard window which is somehow pessimistic e.g. the earlier in 
the prediction window, the heavier is the error. Its definition is the following : 
\[
\begin{aligned}
\delta_{H} T = max\{k \geq 0 \arrowvert \frac{1}{k}\sum_{i=0}^{k}{\frac{|S_{T}(t_{T+i}) - y_{T+i}|}{(y_{T+i}+1)}} \\
\leq 0.05 < \frac{1}{k+1}\sum_{i=0}^{k+1}{\frac{|S_{T}(t_{T+i}) - y_{T+i}|}{(y_{T+i}+1)}} \}
\end{aligned}
\]
For each window type (soft and hard) we normalise it by the size of the observed window $T$.
Results are given for $7$ days, $15$ days and $30$ days scenarios in Table~\ref{table_7d}, Table~\ref{table_15d} and Table~\ref{table_30d} respectively.
In each Table, we give fraction of videos for each model, mean for size of prediction windows (hard and soft) and fraction of videos which meet the bound effect 
(e.g. when the prediction window size is bounded by the horizon). 

\begin{table}[ht!]
 \tiny
 \centering
 \caption{7 days scenario} \label{table_7d}
 \begin{tabular}{|c|c|c|c|c|c|} \hline
  \textbf{Model Type}	& \textbf{Distribution (\%)}	& \textbf{Hard window}	  & \textbf{Hard bounded (\%)}	& \textbf{Soft window}	& \textbf{Soft bounded (\%)} \\ \hline
  E			& $66.9$		& m: $0.6$ 	  & $13.5$			& m: $0.66$  	& $15.3$		\\ \hline 
  ME			& $0.9$			& m: $0.75$ 	  & $17.9$			& m: $0.82$ 	& $20$			\\ \hline
  G			& $10.5$		& m: $0.42$ 	  & $7.8$			& m: $0.47$ 	& $8.8$			\\ \hline 
  MG			& $7.9$			& m: $0.54$ 	  & $10.9$			& m: $0.61$ 	& $12.7$		\\ \hline
  S			& $11.2$		& m: $0.97$ 	  & $32.9$ 			& m: $1   $ 	& $33.8$		\\ \hline
  MS			& $2.6$			& m: $0.81$ 	  & $18.8$			& m: $0.84$ 	& $19.5$		\\ \hline
  All			& $100$			& m: $0.63$ 	  & $15.1$			& m: $0.68$ 	& $16.6$		\\ \hline
 \end{tabular}
\end{table}
%

\begin{table}[ht!]
 \tiny
 \centering
 \caption{15 days scenario} \label{table_15d}
 \begin{tabular}{|c|c|c|c|c|c|} \hline
  \textbf{Model Type}	& \textbf{Distribution (\%)}	& \textbf{Hard window}	  & \textbf{Hard bounded (\%)}	& \textbf{Soft window}	& \textbf{Soft bounded (\%)} \\ \hline
  E			& $62.7$		& m: $0.55$ 	  	  & $10.7$			& m: $0.59$ 	& $12.1$	\\ \hline 
  ME			& $1.3$			& m: $0.9 $ 		  & $22.2$			& m: $0.94$ 	& $23.3$	\\ \hline
  G			& $8.4$			& m: $0.44$ 		  & $7.5$			& m: $0.47$ 	& $7.9$		\\ \hline 
  MG			& $16.5$		& m: $0.7 $ 		  & $13.4$			& m: $0.77$ 	& $15.6$	\\ \hline
  S			& $8.1$			& m: $1.1 $ 		  & $38.1$ 			& m: $1.11$ 	& $38.9$	\\ \hline
  MS			& $3$			& m: $0.94$ 	  	  & $23  $			& m: $0.98$ 	& $24.5$	\\ \hline
  All			& $100$			& m: $0.63$ 		  & $13.6$			& m: $0.67$ 	& $15  $	\\ \hline
 \end{tabular}
\end{table}
%

\begin{table}[ht!]
 \tiny
 \centering
 \caption{30 days scenario} \label{table_30d}
 \begin{tabular}{|c|c|c|c|c|c|} \hline
  \textbf{Model Type}	& \textbf{Distribution (\%)}	& \textbf{Hard window}	  & \textbf{Hard bounded (\%)}	& \textbf{Soft window}	& \textbf{Soft bounded (\%)} \\ \hline
  E			& $52.3$		& m: $0.5$ 		  & $9.5 $			& m: $0.54$  	& $10.5$	\\ \hline 
  ME			& $2.2$			& m: $0.79$ 		  & $17.3$			& m: $0.87$ 	& $19.8$	\\ \hline
  G			& $5.2$			& m: $0.45$ 		  & $8.5$			& m: $0.48$ 	& $9.1$		\\ \hline 
  MG			& $30.6$		& m: $0.68$ 		  & $11.6$			& m: $0.76$ 	& $14  $	\\ \hline
  S			& $5.8$			& m: $1.1 $ 		  & $39.9$ 			& m: $1.14$ 	& $40.6$	\\ \hline
  MS			& $3.9$			& m: $0.89$ 		  & $20.8$			& m: $0.92$ 	& $21.7$	\\ \hline
  All			& $100$			& m: $0.61$ 		  & $12.5$			& m: $0.66$ 	& $13.9$	\\ \hline
 \end{tabular}
\end{table}
%

\section{Conclusion and Future Work}

In the present work we have focused on a method for classifying view-counts dynamics of videos on YouTube. 
We presented different models for YouTube view-count evolution which are able to capture virality and potential population growth. Based on these models,
we have developed one system for automatic classification of the YouTube videos. It aims at classify each YouTube content within one of the four categories we defined: 
Viral and fixed population; Viral and growing population; non-viral fixed population; and non-viral growing population.
We have tested this automatic classification in a particular dataset -that has been presented and is available upon request-.
Results of our experiments reveal that a reasonably small threshold of the $MER$ criterion allows to classify more than $90$\% of the dataset, meaning that the 
defined models explain the  observed behaviour in most of the cases.

Our future work would focus on the context of the four defined categories.
We will analyse how other features influence the dynamic of the view-count evolution.

\bibliographystyle{abbrv}

\bibliography{CR_Bib}

\end{document}